\documentclass[10pt,oneside,final]{IEEEtran}

\usepackage{amssymb}
\usepackage{amsmath}

\usepackage{graphicx} 
\usepackage{subcaption} 
\usepackage{color}
\usepackage{multicol}

\usepackage{bm}
\usepackage{latexsym}
\usepackage{stfloats}
\usepackage{float}
\usepackage{exscale}
\usepackage{relsize}
\usepackage{cite}
\usepackage{cases}
\usepackage{algorithm}
\usepackage{algorithmicx}
\usepackage{algpseudocode}

\usepackage{epsfig}
\usepackage{epstopdf}
\usepackage{breqn}
\usepackage{enumerate}
\usepackage{multirow}
\usepackage[utf8]{inputenc}

\usepackage{mathtools}

\usepackage{setspace}
\usepackage{url}
\usepackage{makecell}
\usepackage{diagbox}
\usepackage{stmaryrd}

\usepackage{xcolor}
\usepackage{xpatch}

\usepackage[colorlinks,
linkcolor=blue,
urlcolor=black,
anchorcolor=blue,
citecolor=blue]{hyperref}

\usepackage{accents} 

\algrenewcommand{\algorithmicrequire}{\textbf{Input:}}  
\algrenewcommand{\algorithmicensure}{\textbf{Output:}} 

\begin{document}

	\title{\vspace{-0.4cm}Cooperative Bistatic ISAC Systems for Low-Altitude Economy\vspace{-0.1cm}}
	
	\author{
		Zhenkun Zhang, 
		Yining Xu, 
		Cunhua Pan, 
		Hong Ren,
		Qixuan Zhang,
		Songtao Gao,
		
		and Jiangzhou Wang, \textit{Fellow, IEEE}
		\thanks{
			Zhenkun Zhang,	Yining Xu, Cunhua Pan, Hong Ren and Jiangzhou Wang are with National Mobile Communications Research Laboratory, Southeast University, Nanjing 210096, China (e-mail: \{zhenkun\_zhang, yining.xu, cpan, hren, j.z.wang\}@seu.edu.cn).
			Qixuan Zhang and Songtao Gao are with China Mobile Group Design Institute Co., Ltd., Beijing, China. (e-mail: \{zhangqixuan, gaosongtao\}@cmdi.chinamobile.com).
			
			Zhenkun Zhang and Yining Xu are co-first authors.
			
			\textit{Corresponding author: Cunhua pan}}
		\vspace{-0.9cm}
	}
	%
	
	%
		%
	\maketitle
	
	\newtheorem{lemma}{Lemma}
	\newtheorem{theorem}{Theorem}
	\newtheorem{remark}{Remark}
	\newtheorem{corollary}{Corollary}
	\newtheorem{proposition}{Proposition}
	
	\newcommand{\mathcalbf}[1]{\bm{\mathcal{#1}}}

	%
	\begin{abstract}
		The burgeoning low-altitude economy (LAE) necessitates integrated sensing and communication (ISAC) systems capable of high-accuracy multi-target localization and velocity estimation under hardware and coverage constraints inherent in conventional ISAC architectures. 
		This paper addresses these challenges by proposing a cooperative bistatic ISAC framework within MIMO-OFDM cellular networks, enabling robust sensing services for LAE applications through standardized 5G New Radio (NR) infrastructure.
		We first develop a low-complexity parameter extraction algorithm employing CANDECOMP/PARAFAC (CP) tensor decomposition, which exploits the inherent Vandermonde structure in delay-related factor matrices to efficiently recover bistatic ranges, Doppler velocities, and angles-of-arrival (AoA) from multi-dimensional received signal tensors.
		To resolve data association ambiguity across distributed transmitter-receiver pairs and mitigate erroneous estimates, we further design a robust fusion scheme based on the minimum spanning tree (MST) method, enabling joint 3D position and velocity reconstruction.
		Comprehensive simulation results validate the framework’s superiority in computational efficiency and sensing performance for low-altitude scenarios.

		
	\end{abstract}

	%
	
	%
	
	%
	\vspace{-0.3cm}
	\section{Introduction}
	The rapid evolution toward sixth-generation (6G) and NextG wireless networks has ushered in a transformative vision where communication systems transcend traditional data transmission to embrace advanced sensing capabilities \cite{9737357}.
	Integrated sensing and communication (ISAC) is central to realizing this vision, unifying sensing and communication within shared infrastructure.
	This convergence promises significant gains for optimized spectral efficiency, reduced hardware costs, and novel application domains \cite{8999605}.
	By exploiting the inherent synergy between sensing and communication, ISAC not only enhances network adaptability but also paves the way for intelligent environments where real-time situational awareness coexists with ultra-reliable connectivity.
	ISAC is particularly transformative for the low-altitude economy (LAE) and its supporting low-altitude wireless networks (LAWNs), where its capabilities align directly with critical operational needs \cite{9456851}.
	LAE encompasses applications spanning logistics, agriculture, and emergency services, which rely on unmanned aerial vehicles (UAVs) operating in complex airspace, demanding robust detection, localization, and tracking capabilities alongside seamless connectivity.
	Addressing these requirements necessitates ISAC protocols tailored to LAE’s unique challenges, including high-accuracy multi-target sensing and compatibility with the existing cellular infrastructure.
	
	Similar to radar systems, ISAC implementations typically adopt two basic operational modes: \textit{monostatic (a.k.a. active sensing)} \cite{9367457,9724260} and \textit{bistatic (a.k.a. passive sensing)} \cite{9860521}. 
	In monostatic systems, a single base station (BS) simultaneously transmits and receives sensing signals, enabling independent operation.
	However, this approach requires full-duplex transceivers with self-interference mitigation capability, which is not supported by the existing commercial BSs, necessitating costly hardware upgrades.
	Conversely, bistatic configurations separate the transmitter and receiver.
	This sensing pattern broadens coverage and avoids costly full-duplex hardware, facilitating integration with the existing systems \cite{WAN2020,10614082}, while introducing synchronization requirements between BSs.
	In addition to these challenges, conventional configurations-whether monostatic or bistatic-also have inherent limitations: sparse spatial coverage due to limited transmission power and environmental obstacles, speed estimation restricted to the radial component, and susceptibility to measurement outliers.
	
	Cooperative ISAC architectures, inspired by \textit{multistatic} radar principles, are expected to overcome these limitations \cite{10273396,10032141}. 
	By leveraging distributed BSs to collaboratively transmit and receive sensing signals, cooperative ISAC systems amplify coverage, enhance measurement diversity, and improve robustness against environmental dynamics.
	In cooperative ISAC networks, multiple BSs can generate rich spatial-temporal measurements.
	Sensing parameters such as time delays, Doppler shifts, and angles are extracted locally at BSs or centrally at a central processing unit (CPU).
	These parameters are then fused at the CPU for high-accuracy target localization and true velocity reconstruction. 
	In our previous work \cite{zhenkunzhang}, a cell-free-based cooperative bistatic ISAC framework was proposed.
	It revealed that such cooperation can be achieved through protocol-level innovations, such as staggered uplink/downlink scheduling across BSs, which repurposed existing orthogonal frequency-division multiplexing (OFDM) communication waveforms and infrastructure for cooperative sensing without hardware updates.
	This alignment with the standardized 5G New Radio (NR) frame structure demonstrated how cooperative ISAC combines theoretical advancements with practical deployment capabilities, which is a key consideration for 6G standardization.
	
	Prior research has established foundational frameworks for cooperative ISAC-enabled sensing schemes.
	For example, \cite{zhenkunzhang} and \cite{9724258} addressed single-antenna BS configurations, employing bistatic and monostatic sensing modes through two-dimensional (2D) fast Fourier transform (2D-FFT) and compressed sensing algorithms for target localization.
	Subsequent studies extended these concepts to multi-antenna architectures. 
	For the scenario with multi-antenna BSs, Lu et al. \cite{10615952} investigated a multi-BS cooperative system using multiple signal classification (MUSIC) algorithms for parameter estimation, complemented by a lattice point search method for data fusion. 
	This progression continued with Wei et al. \cite{10226276}, who developed a symbol-level data fusion scheme integrating phase features in the demodulation symbols with distance and radial velocity estimations in cooperative monostatic ISAC systems, later adapted by the same research group \cite{10787076} for bistatic configurations with enhanced position and velocity estimation capabilities based on angle-of-arrival (AoA) and angle-of-departure (AoD) extractions.
	In addition, feasible topology and frame structure designs were proposed by Han et al. \cite{Han2024}, which achieved cooperative bistatic sensing in the cellular networks at a low cost.
	A hybrid monostatic-bistatic architecture was proposed by Jiang et al. \cite{10616023}, where dual BSs jointly process echo signals from both BSs for improved target localization.
	Emerging solutions attempted to address constraints in simplified scenarios through advanced mathematical tools.
	Our previous work \cite{JunTang} introduced tensor decomposition techniques to cooperative monostatic systems, enabling low-complexity parameter estimation of range, angle, and velocity with inherent multi-dimensional parameter pairing.
	This approach has already been used in channel estimation \cite{zhouzhou,11069254}, showing promise for the ISAC-enabled sensing scheme designs \cite{ruoyu_zhang}.
	
	Given the extensive coverage of mobile networks, developing sensing schemes compatible with existing infrastructure is key for practical ISAC deployment and application spread. 
	However, despite the prevalence of massive MIMO BSs, existing studies mainly addressed simplified scenarios, such as single-antenna configurations \cite{9724258,zhenkunzhang}, single-target sensing \cite{10615952,10226276,10787076,Han2024}, and 2D spatial estimation \cite{9724258,zhenkunzhang,10226276,10787076,10616023,Han2024}, failing to fully leverage network capabilities or reflect the complexity of practical deployments like LAE.
	In this context, bistatic architectures \cite{zhenkunzhang,10787076,Han2024} have inherent advantages over monostatic approaches \cite{9724258,10226276,10615952,JunTang} in hardware compatibility by reusing the existing communication equipment, while also generating richer sensing data. 
	For instance, with 5 cooperative BSs configured as 2 transmitters and 3 receivers, bistatic schemes can generate $2\times 3$ measurement links, exceeding the 5 self-transceiving links in monostatic schemes.
	Meanwhile, fewer transmitters reduce spectrum resource consumption and inter-BS interference.
	Nevertheless, fusing abundant bistatic measurements to enhance sensing accuracy remains challenging.
	While \cite{zhenkunzhang} proposed a solution for single-antenna scenarios, the high complexity limits its scalability in environments with a large number of BSs or targets.
	Although the data fusion framework in \cite{JunTang} provided a viable paradigm for monostatic systems, it cannot be directly applied to bistatic architectures due to inherent challenges in bistatic sensing.
	For example, symbol timing offset (STO) and carrier frequency offset (CFO) need to be effectively suppressed, otherwise serious range and velocity estimation errors will occur.
	In addition, due to the geographical separation between transmitters and receivers in bistatic sensing, the angle of departure (AOD) and angle of arrival (AOA) related to a target are different, and the distance between a target and a BS cannot be directly estimated, which increases the difficulty of parameter estimation and data fusion.
	
	
	This paper addresses the aforementioned gaps by developing a cooperative bistatic ISAC framework for LAE scenarios.
	Within a MIMO-OFDM cellular network, we introduce a bistatic sensing architecture compatible with standard 5G NR frame structures.
	To achieve high-accuracy position and velocity estimation, efficient signal processing methods are designed for sensing parameter extraction and data fusion.
	The main contributions of this work are summarized as follows:
	\begin{enumerate}
		\item 
		We propose a general cooperative bistatic ISAC framework based on MIMO-OFDM cellular networks that enable multi-BS multi-target cooperative sensing with minimum hardware modifications, maintaining compatibility with 5G NR frame structures.
		
		\item 
		We develop a low-complexity parameter estimation algorithm using Estimation of Signal Parameters via Rotational Invariance Techniques (ESPRIT)-inspired CANDECOMP/PARAFAC (CP) tensor decomposition, which efficiently extracts range, velocity, and angle parameters of the targets by leveraging the Vandermonde structure of delay-related factor matrices.
		
		\item 
		We design a robust data fusion scheme that employs a minimum spanning tree (MST)-based association method to eliminate erroneous estimates and resolve data association ambiguity among transmitter-receiver pairs, thereby enabling joint 3D position and velocity estimation.
		
		\item 
		Comprehensive simulation results validate the superiority of the proposed framework over the baseline schemes in terms of computational complexity and sensing performance in LAE scenarios.
		
	\end{enumerate}
	
	The remainder of this paper is organized as follows. 
	Section \ref{Sec_SystemModel} introduces the system model and bistatic channel formulation. 
	Section \ref{Sec_Tensor-Based_Estimation} details the tensor-based parameter extraction algorithm with uniqueness analysis. 
	Section \ref{Sec_Pos_Vel_Est} presents the MST-based data fusion framework.
	Section \ref{Sec_Simu_Result} evaluates the performance of the proposed schemes through simulations, and Section \ref{Sec_Conclusion} concludes the work.

	\emph{Notations}:
	Mathematical entities are typographically distinguished as follows: 
	scalars use lowercase letters ($y$), vectors bold lowercase ($\mathbf{y}$), matrices bold uppercase ($\mathbf{Y}$), sets calligraphic uppercase ($\mathcal{Y}$), and tensors bold calligraphic uppercase ($\mathcalbf{Y}$). 
	The real and complex number fields are denoted by $\mathbb{R}$ and $\mathbb{C}$, respectively.
	Fundamental operators are defined as: $\circ$ for outer product, $\otimes$ for Kronecker product, $*$ for Hadamard product, and $\odot$ for Khatri-Rao product.
	The transpose, conjugate transpose, inverse, and Moore-Penrose pseudo-inverse of a matrix $\mathbf{Y}$ are denoted by $\mathbf{Y}^{\mathrm{T}}$, $\mathbf{Y}^{\mathrm{H}}$, $\mathbf{Y}^{-1}$, and $\mathbf{Y}^{\dagger}$, respectively. 
	The $\ell_1$ norm and $\ell_2$ norm of a vector $\mathbf{y}$ are expressed as $\left\| \mathbf{y} \right\|_1$ and $\left\| \mathbf{y} \right\|_2$, respectively. 
	The operator $\mathrm{diag}\left( \mathbf{y} \right) $ generates a diagonal matrix from vector $\mathbf{y}$, while $\text{unvec}_{M \times N}\left( \cdot \right)$ reshapes an $M N \times 1$ vector $\mathbf{y}$ into an $M \times N$ matrix $\mathbf{Y}$.

	\vspace{-0.1cm}
	\section{System Model}\label{Sec_SystemModel}
	\subsection{Signal Transmission Model}
\begin{figure}[!tb]
		\centering
		\includegraphics[width=0.8\linewidth]{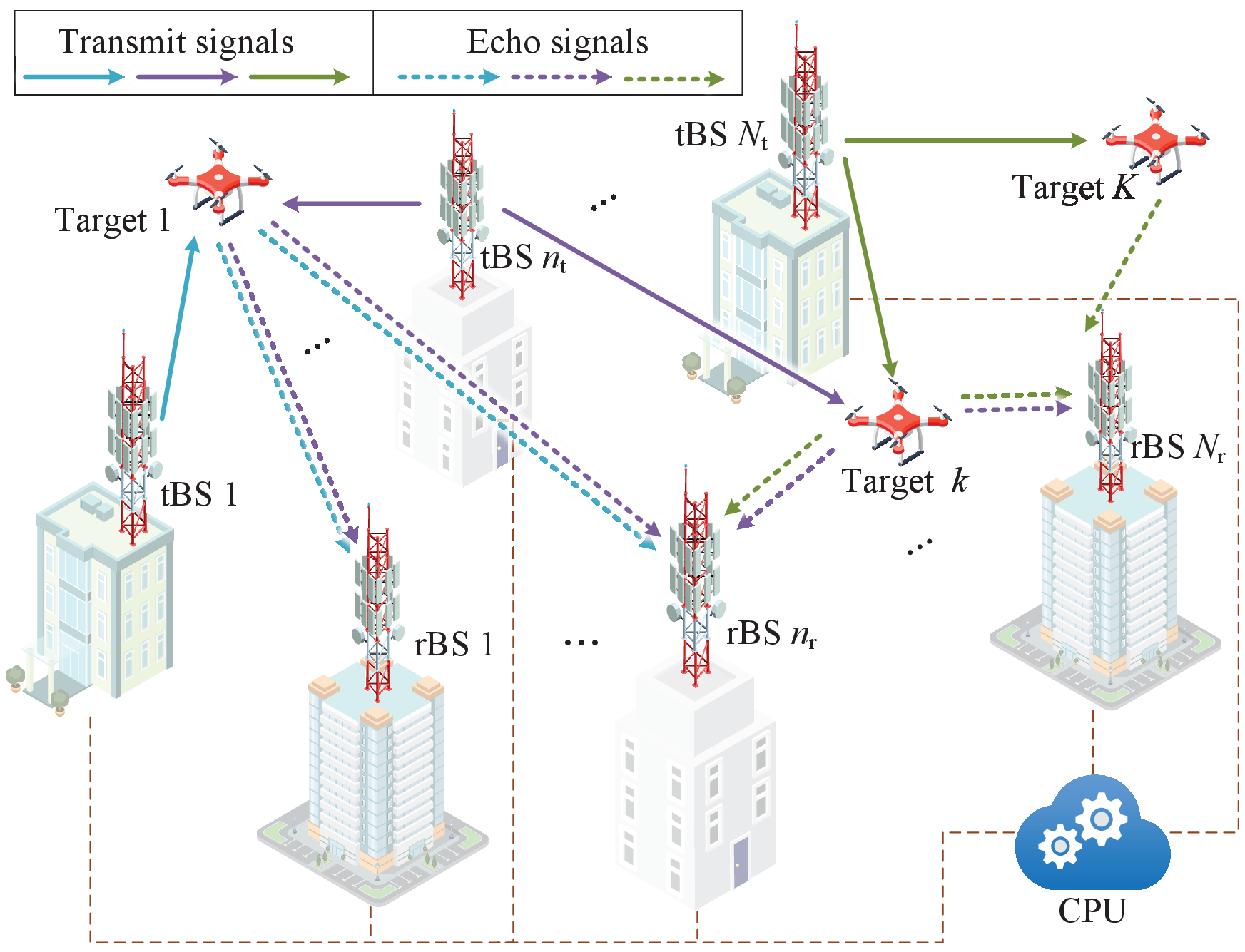}
		\caption{An illustration of the MIMO-OFDM cellular network-based bistatic ISAC architecture. }
		\label{fig:system_model}
		\vspace{-0.1cm}
	\end{figure}

\begin{figure}[!t]
		\centering
		\includegraphics[width=0.9\linewidth]{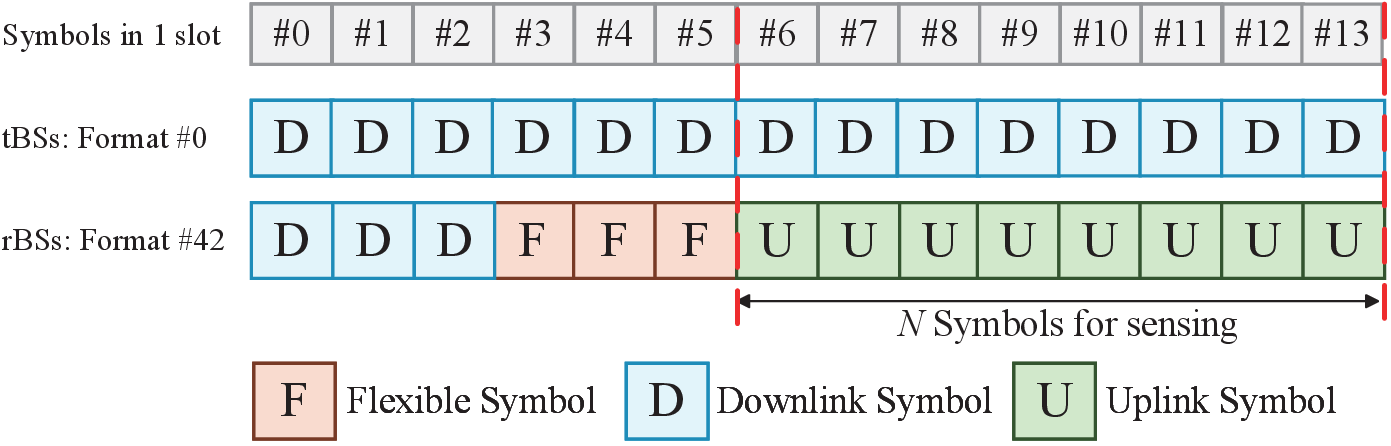}
		\caption{An example of feasible slot format configuration, where a duration of $N=8$ symbols can be used for sensing signal.}
		\label{fig:frame}
		\vspace{-0.2cm}
	\end{figure}

	As illustrated in Fig. \ref{fig:system_model}, we consider a MIMO-OFDM network-based cooperative ISAC system capable of detecting, locating, and measuring velocities of low-altitude targets, thereby supporting the low-altitude economy applications.
	This system employs a bistatic sensing architecture, where $N_{\mathrm{t}}$ transmit BSs (tBSs) simultaneously emit sensing signals, while $N_{\mathrm{r}}$ receive BSs (rBSs) collaboratively estimate the parameters of $K$ targets through processing the reflected echoes.\footnote{
		While ISAC systems can dynamically reconfigure tBS/rBS assignments based on specific sensing requirements, this work focuses exclusively on developing sensing algorithms, with BS selection strategies proposed in \cite{2024arXiv241220349R}.
	}
	Operating in the time division duplexing (TDD) mode, all BSs are connected with a CPU through fronthaul links.
	The CPU sends timing information to the BSs via the fronthaul links for synchronization, performs data fusion of the rBS-derived sensing information to enhance the sensing performance, and adjusts the slot format \cite{3GPP138213} of each BS to facilitate the sensing signal exchange.
	A detailed example of the slot format configuration is presented in Fig. \ref{fig:frame}.
	Similar to the cooperative ISAC framework established in our previous work \cite{zhenkunzhang}, to achieve bistatic sensing in mobile communication systems, the rBSs are configured with slot formats different from the tBSs when performing sensing tasks.
	During the $N$-symbol sensing task, each tBS transmits pre-determined sensing reference signals on non-overlapping bandwidth parts (BWPs) to eliminate the inter-BS interference.\footnote{To minimize modifications to existing transmission protocols, dedicated symbol resources are allocated for sensing by suspending communication services. 
		Although this incurs throughput loss, the impact is negligible since sensing tasks require only a few symbols and occur at relatively low refresh rates (e.g., once every few seconds or a few times per second).}
	The estimated parameters from these signals are then uploaded to the CPU through fronthaul links.
	
	Within the common 5G telecommunication infrastructure, every BS implements a MIMO-OFDM air interface employing uniform planar arrays (UPAs) for 3D beamforming \cite{3GPP138104}.
	Specifically, the UPAs are installed with an elevation angle of 0° to facilitate both the sensing of low-altitude targets and the communication service for terrestrial users.
	To achieve cost-efficient massive antenna deployment, we consider an architecture that all BSs adopt the fully-connected hybrid beamforming structure, which reduce the number of radio frequency (RF) chains. 
	We assume that each BS comprises $R$ RF chains driving $L = N_{\mathrm{h}} \times N_{\mathrm{v}} > R$ antenna elements, where $N_{\mathrm{v}}$ and $N_{\mathrm{h}}$ denote the numbers of antenna elements in the vertical and horizontal directions, respectively.
	
	Note that each rBS can effectively separate and identify the transmitted signals from each tBS by applying bandpass filtering to each BWP.
	In addition, all rBSs in the considered sensing framework independently execute identical processing algorithms on signals transmitted from individual tBSs. 
	Therefore, we consider the signal transmitted from tBS $n_{\mathrm{t}}$ to rBS $n_{\mathrm{r}}$ as an illustrative case for presenting the transmission model and the proposed parameter extraction scheme. 
	For notational compactness, we omit the subscripts $n_{\mathrm{t}}$ and $n_{\mathrm{r}}$ throughout the derivations in the following.
	Let $s_{r,m,n} $ denote the modulated frequency-domain sensing symbol allocated in the $m$th subcarrier of the $n$th OFDM symbol input to the $r$th RF chain, which satisfies $\left| s_{r,m,n} \right|^2=1$.
	Then, the baseband time-domain OFDM signal transmitted by tBS $n_{\mathrm{t}}$ can be represented as 
	\begin{equation}
		s_r\left( t \right) =\sum_{m=1}^M{\sum_{n=1}^N{s_{r,m,n}}e^{j2\pi m\Delta ft}}\cdot r\left( t-nT_{\mathrm{s}} \right) ,\, l=1,...,L,
	\end{equation}
	where $M$ is the number of subcarriers, $T_{\mathrm{s}}$ denotes the OFDM symbol period (including the cyclic prefix), $r(t)$ denotes the pulse-shaping filter function, and $\Delta f$ denotes the subcarrier spacing (SCS). 
	Following matched filtering, cyclic prefix removing, and inverse fast Fourier transform (IFFT), the frequency-domain signal vector received at the rBS during the $n$th sensing symbol is expressed as
	\begin{align}
		\mathbf{y}_{m,n}=\mathbf{Q}^{\mathrm{H}}\mathbf{H}_{m,n}\mathbf{Fs}_{m,n}+\mathbf{Q}^{\mathrm{H}}\mathbf{n}_{m,n}\in \mathbb{C} ^{R\times 1},
	\end{align}
	where $\mathbf{H}_{m,n}\in \mathbb{C}^{L\times L}$ denotes the discrete frequency-domain channel (its formulation will be derived in the subsequent subsection),
	the hybrid precoding matrix $\mathbf{F} \triangleq \mathbf{F}_{\mathrm{a}}\mathbf{F}_{\mathrm{d}}$ combines an analog phase-shifting network $\mathbf{F}_{\mathrm{a}} \in \mathbb{C}^{L\times R}$ with a digital precoder $\mathbf{F}_{\mathrm{d}} \in \mathbb{C}^{R\times R}$, 
	while the hybrid combining matrix $\mathbf{Q} = \mathbf{Q}_{\mathrm{a}}\mathbf{Q}_{\mathrm{d}}$ integrates an analog combining matrix $\mathbf{Q}_{\mathrm{a}} \in \mathbb{C}^{L\times R}$ and a digital combining matrix $\mathbf{Q}_{\mathrm{d}} \in \mathbb{C}^{R\times R}$.
	The transmitted symbol vector $\mathbf{s}_{m,n}=[s_{1,m,n},\cdots ,s_{r,m,n},\cdots ,s_{R,m,n}]^{\mathrm{T}}\in \mathbb{C} ^{L\times 1}$ contains the sensing symbols from tBS $n_{\mathrm{t}}$, and $\mathbf{n}_{m,n} \in \mathbb{C}^{L \times 1}$ represents the additive white Gaussian noise vector.
	Given the known symbol vector $\mathbf{s}_{m,n}$, the rBSs can eliminate the impact of it by multiplying the received signal vector by $\mathbf{s}_{m,n}^{*}$, i.e.,
	\begin{align}
		\hat{\mathbf{y}}_{m,n}= \mathbf{y}_{m,n}*\mathbf{s}_{m,n}^{*} =\mathbf{Q}^{\mathrm{H}}\mathbf{H}_{m,n}\mathbf{F1}+ \hat{\mathbf{n}}_{m,n},
		\label{equation:hat_y}
	\end{align}
	where 
	$\mathbf{1}$ is an all-one column vector and $\hat{\mathbf{n}}_{m,n} 
	\in\mathbb{C}^{L \times 1}$ is the equivalent noise vector.
	
	\vspace{-0.1cm}
	\subsection{Beanforming for Low-altitude Targets} \label{Sec_Beanforming_LAE}
\begin{figure}[tbp]
		\centering
		\begin{subfigure}[b]{0.45\linewidth}
			\centering
			\includegraphics[width=\linewidth]{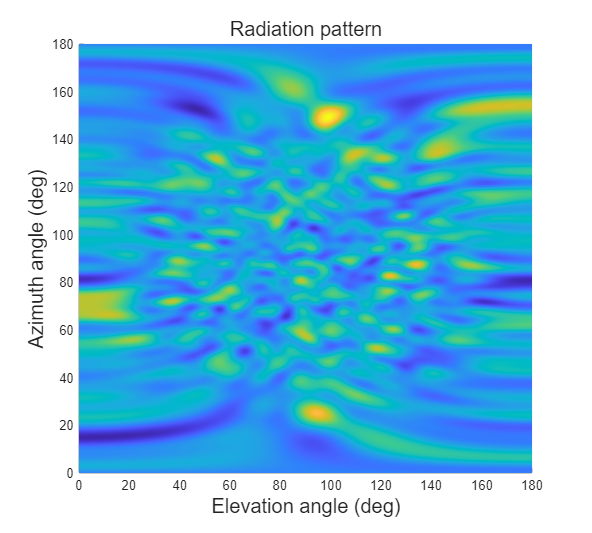}
			\caption{Random}
			\label{fig:radiation_pattern_random}
		\end{subfigure}
		\hfill
		\begin{subfigure}[b]{0.45\linewidth}
			\centering
			\includegraphics[width=\linewidth]{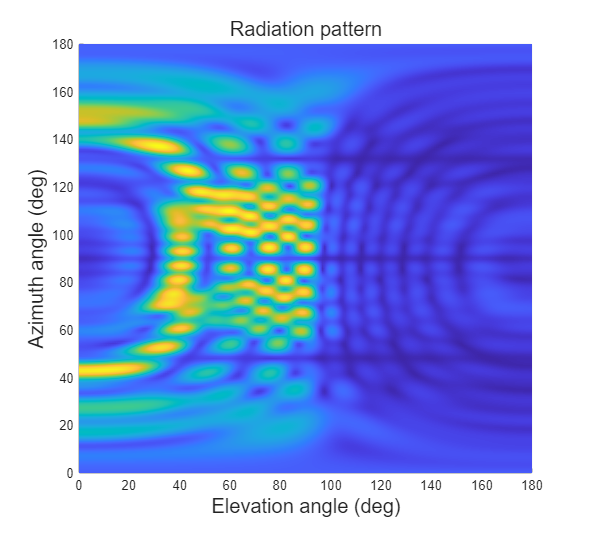}
			\caption{Designed}
			\label{fig:radiation_pattern_designed}
		\end{subfigure}
		\caption{Radiation patterns achieved by random beamforming and the exemplary design for low-altitude target sensing.}
		\label{fig:radiation_pattern}
		\vspace{-0.4cm}
	\end{figure}

	In practical sensing scenarios, the positions of low-altitude targets are typically unknown a priori. 
	An Intuitive method to enhance the echo signal-to-noise ratio (SNR) and improve sensing accuracy is to perform beam scanning to obtain coarse directional information of the targets, followed by steering beams toward these directions \cite{JunTang}. 
	However, due to the narrow beamwidth, exhaustive beam scanning using large antenna arrays incurs a substantial time overhead.
	Conversely, random beamforming tends to spread transmit energy excessively across space, as illustrated in Fig. \ref{fig:radiation_pattern_random}, leading to weak target echoes and strong ground clutter.
	To balance beamforming gain and clutter suppression, an appropriate design of the beamforming/combining matrices $\mathbf{F}$ and $\mathbf{Q}$ is essential. 
	Based on the above discussions, an ideal beamforming scheme should concentrate radiated energy within the sensing region of interest. 
	Specifically, let $\mathcal{R}$ denote the angular region defined by elevation range $[\theta_{\min}, \theta_{\max}]$ and azimuth range $[\phi_{\min}, \phi_{\max}]$. 
	The beamforming gain $\mathrm{Gain}(\theta, \phi)$ should satisfy:
	\begin{equation}
		\mathrm{Gain}\left( \theta ,\phi \right) \approx \begin{cases}
			\mathrm{Gain}_0, \left( \theta ,\phi \right) \in \mathcal{R},\\
			\varepsilon _0, \text{otherwise},\\
		\end{cases}
	\end{equation}
	where $\varepsilon_0 \ll \mathrm{Gain}_0$ to suppress ground clutter.
	That is, the gain should be approximately uniform and high within $\mathcal{R}$, and minimized outside it.
	Although precoding design is a well-studied topic and not the focus of this work, we provide an exemplary design for the precoding matrix $\mathbf{F}$ to illustrate that such beam patterns are feasible in practice. 
	A similar approach can be applied to construct the combining matrix $\mathbf{Q}$.
	
	Consider a sensing region defined by elevation range from 40° to 90° and azimuth range from 40° to 140°. 
	With $R = 64$ RF chains, we uniformly sample this region into 64 angle pairs, using 8 elevation samples $\left[ \theta_{1}^{\mathrm{samp}}, \dots, \theta_{8}^{\mathrm{samp}} \right] = \left[ 40, 47.14, \dots, 90 \right]$ and 8 azimuth samples $\left[ \phi_{1}^{\mathrm{samp}}, \dots, \phi_{8}^{\mathrm{samp}} \right] = \left[ 40, 54.29, \dots, 140 \right]$. 
	Let $\mathbf{F}_{\mathrm{d}} = \mathbf{I}$ and construct $\mathbf{F}_{\mathrm{a}}$ such that its $\left( p_{\mathrm{s}} + 8(q_{\mathrm{s}} - 1) \right)$-th column is given by the transmit steering vector $\mathbf{a}_{\mathrm{t}}\left( \theta _{p_{\mathrm{s}}}^{\mathrm{samp}},\phi _{q_{\mathrm{s}}}^{\mathrm{samp}} \right) $, whose definition will be given in the next subsection. 
	As shown in Fig. \ref{fig:radiation_pattern_designed}, this design effectively suppresses energy leakage toward the ground.
	
	It is worth emphasizing that the above example is neither the only feasible design nor necessarily optimal. 
	Nevertheless, it demonstrates that by properly configuring beamforming matrices $\mathbf{F}$ and $\mathbf{Q}$ at the tBSs and rBSs, ground clutter can be significantly suppressed. 
	Therefore, in the subsequent derivations of this paper, we ignore the ground clutter component in the channel $\mathbf{H}_{m,n}$.

	\subsection{Sensing Channel Model}
	In low-altitude sensing scenarios, consistent with common assumptions in cooperative ISAC systems \cite{zhenkunzhang,JunTang}, we assume a direct line-of-sight (LoS) path between any tBS-rBS pair. 
	This leads to $K+1$ resolvable propagation paths at rBS $n_{\mathrm{r}}$, comprising $K$ target-scattered paths and one dominant LoS component, where the length of target-scattered paths are also known as the \textit{bistatic ranges}.
	The temporal characteristics of these paths are quantified through path-specific delays $\tau_k=d_k/c_0$ for $k=1,\dots,K+1$, where $d_k$ represents the length of the $k$th path, and $c_0$ is the speed of light.
	In addition, the Doppler shift of the LoS path ($k=k_{\mathrm{LoS}}$) is known to satisfy $f_{k_{\mathrm{LoS}}}^{\mathrm{d}}=0$, while each scattered path ($k\ne k_{\mathrm{LoS}}$) exhibits shift $f_{k}^{\mathrm{d}}=v_k/\lambda$ proportional to the total radial velocity $v_k$ (called \textit{bistatic Doppler velocity}) of a target relative to tBS $n_{\mathrm{t}}$ and rBS $n_{\mathrm{r}}$, with $\lambda$ denoting the carrier wavelength.
	Then, the time-delay domain channel matrix between tBS $n_{\mathrm{t}}$ and rBS $n_{\mathrm{r}}$ is formulated as
	\begin{align}
		\label{equation:time_and_delay_channel}
		\mathbf{H}(t,\tau)=\sum_{k=1}^{K+1}{\left[ \begin{array}{l}
				\alpha _k\mathbf{a}_{\mathrm{r}}\left( \theta _{k}^{\mathrm{r}},\phi _{k}^{\mathrm{r}} \right) \mathbf{a}_{\mathrm{t}}^{\mathrm{H}}\left( \theta _{k}^{\mathrm{t}},\phi _{k}^{\mathrm{t}} \right)\\
				\times \delta \left( \tau -\tau _k-{\delta}^{\mathrm{t}} \right) e^{j2\pi (f_{k}^{\mathrm{d}}+{\delta}^{\mathrm{f}})t}\\
			\end{array} \right] ,}
	\end{align}
	where $\alpha_k$ represents the composite channel gain of the $k$th path, while ${\delta}^{\mathrm{t}}$ and ${\delta}^{\mathrm{f}}$ are the STO and CFO after synchronization between tBS $n_{\mathrm{t}}$ and rBS $n_{\mathrm{r}}$.
	The channel gain is influenced by factors such as the path length $d_k$ and radar cross-section (RCS) $\sigma _{\mathrm{RCS}}$ of targets, making direct estimation challenging. 
	The spatial characteristics are determined by elevation/azimuth AoA $\left\{ \theta _{k}^{\mathrm{r}},\phi _{k}^{\mathrm{r}} \right\} $ and AoD $\left\{ \theta _{k}^{\mathrm{t}},\phi _{k}^{\mathrm{t}} \right\} $.
	To simplify the notation, we further define
	\begin{equation}
		\left\{ \begin{array}{l}
			\varPhi _{k}^{\mathrm{r}}\triangleq \cos \theta _{k}^{\mathrm{r}}, \,\varTheta _{k}^{\mathrm{r}}\triangleq \sin \theta _{k}^{\mathrm{r}}\cos \phi _{k}^{\mathrm{r}},\\
			\varPhi _{k}^{\mathrm{t}}\triangleq \cos \theta _{k}^{\mathrm{t}}, \,\varTheta _{k}^{\mathrm{t}}\triangleq \sin \theta _{k}^{\mathrm{t}}\cos \phi _{k}^{\mathrm{t}}.\\
		\end{array} \right.
		\label{equation:azimuth/elevationAoA}
	\end{equation}
	Then, the steering vectors $\mathbf{a}_{\mathrm{r}}\left( \theta _{k}^{\mathrm{r}},\phi _{k}^{\mathrm{r}} \right)$ and $\mathbf{a}_{\mathrm{t}}\left( \theta _{k}^{\mathrm{t}},\phi _{k}^{\mathrm{t}} \right) $ in \eqref{equation:time_and_delay_channel} can be written as
	\begin{align}
		&\mathbf{a}_{\mathrm{r}}\left( \varTheta _{k}^{\mathrm{r}},\varPhi _{k}^{\mathrm{r}} \right) =\mathbf{a}_{\mathrm{r}}^{\mathrm{v}}\left( \varPhi _{k}^{\mathrm{r}} \right) \otimes \mathbf{a}_{\mathrm{r}}^{\mathrm{h}}\left( \varTheta _{k}^{\mathrm{r}} \right) \in \mathbb{C} ^{L\times 1},
		\\
		&\mathbf{a}_{\mathrm{t}}\left( \varTheta _{k}^{\mathrm{t}},\varPhi _{k}^{\mathrm{t}} \right) =\mathbf{a}_{\mathrm{t}}^{\mathrm{v}}\left( \varPhi _{k}^{\mathrm{t}} \right) \otimes \mathrm{}\mathbf{a}_{\mathrm{t}}^{\mathrm{h}}\left( \varTheta _{k}^{\mathrm{t}} \right) \in \mathbb{C} ^{L\times 1},
	\end{align}
	where the corresponding UPA steering vectors are constructed based on planar wavefront geometry
	\begin{align}
		&\mathbf{a}_{\mathrm{r}}^{\mathrm{v}}\left( \varPhi _{k}^{\mathrm{r}} \right) =[1,\cdots ,e^{j2\pi (N_{\mathrm{v}}-1)\frac{d}{\lambda}\varPhi _{k}^{\mathrm{r}}}]^{\mathrm{T}}\in \mathbb{C} ^{N_{\mathrm{v}}\times 1},
		\\
		&\mathbf{a}_{\mathrm{r}}^{\mathrm{h}}\left( \varTheta _{k}^{\mathrm{r}} \right) =[1,\cdots ,e^{j2\pi (N_{\mathrm{h}}-1)\frac{d}{\lambda}\varTheta _{k}^{\mathrm{r}}}]^{\mathrm{T}}\in \mathbb{C} ^{N_{\mathrm{h}}\times 1},
	\end{align}
	with analogous expressions for $\mathbf{a}_{\mathrm{t}}^{\mathrm{v}}\left( \varPhi _{k}^{\mathrm{t}} \right) $ and $\mathbf{a}_{\mathrm{t}}^{\mathrm{h}}\left( \varTheta _{k}^{\mathrm{t}} \right)$ through substitution of superscripts $\mathrm{r}\rightarrow \mathrm{t}$.
	By processing the Fourier transform of $\tau$, the frequency domain channel at the $m$th subcarrier can be derived as
	\begin{equation}
		\mathbf{H}_m(t)=\sum_{k=1}^{K+1}{\left[ \begin{array}{l}
				\alpha _k\mathbf{a}_{\mathrm{r}}\left( \varTheta _{k}^{\mathrm{r}},\varPhi _{k}^{\mathrm{r}} \right) \mathbf{a}_{\mathrm{t}}^{\mathrm{H}}\left( \varTheta _{k}^{\mathrm{t}},\varPhi _{k}^{\mathrm{t}} \right)\\
				\times e^{-j2\pi m\Delta f\left( \tau _k+{\delta}^{\mathrm{t}} \right)}e^{j2\pi \left( f_{k}^{\mathrm{d}}+{\delta}^{\mathrm{f}} \right) t}\\
			\end{array} \right] .}
	\end{equation}
	Then, by sampling the received signal at the $n$th OFDM symbol, we can obtain the discrete frequency-domain channel for the $n$th symbol at the $m$th subcarrier as follows
	\begin{equation}\label{equation:channel_H_mn}
		\mathbf{H}_{m,n}=\sum_{k=1}^{K+1}{\left[ \begin{array}{l}
				\alpha _k\mathbf{a}_{\mathrm{r}}\left( \varTheta _{k}^{\mathrm{r}},\varPhi _{k}^{\mathrm{r}} \right) \mathbf{a}_{\mathrm{t}}^{\mathrm{H}}\left( \varTheta _{k}^{\mathrm{t}},\varPhi _{k}^{\mathrm{t}} \right)\\
				\times e^{-j2\pi m\Delta f\left( \tau _k+{\delta}^{\mathrm{t}} \right)}e^{j2\pi \left( f_{k}^{\mathrm{d}}+{\delta}^{\mathrm{f}} \right) nT_{\mathrm{s}}}\\
			\end{array} \right] .}
	\end{equation}

	In the following, we first develop an efficient sensing parameter estimation algorithm to extract the bistatic ranges $\left\{ d_k \right\} _{k=1}^{K+1}$, azimuth/elevation angles $\left\{ \theta _{k}^{\mathrm{r}},\phi _{k}^{\mathrm{r}} \right\} _{k=1}^{K+1}$, and bistatic Doppler velocities $\left\{ v_k \right\} _{k=1}^{K+1}$ related to the targets from the received signals $\hat{\mathbf{y}}_{m,n}$.
	Then, we propose a data fusion scheme that integrates these parameters to achieve accurate 3D localization and velocity estimation of the targets. 
	
	\section{Tensor-Based Parameter Estimation Algorithm}\label{Sec_Tensor-Based_Estimation}
	In this section, a tensor decomposition-based parameter estimation framework for bistatic ISAC systems is proposed. 
	By reshaping received signals into a third-order tensor, we employ CP decomposition to extract the delay, angle, and Doppler factor matrices from the received signals $\hat{\mathbf{y}}_{m,n}$. 
	Leveraging the inherent Vandermonde structure in one factor matrix, we develop an ESPRIT-inspired algorithm for efficient parameter estimation. 
	
	\subsection{Tensor Formulation of the Received Signal}
	By substituting \eqref{equation:channel_H_mn} into \eqref{equation:hat_y}, the received signal $\hat{\mathbf{y}}_{m,n}$ can be rewritten as
	\begin{align}
		\hat{\mathbf{y}}_{m,n}=\sum_{k=1}^{K+1}{\left[ \begin{array}{l}
				\alpha _k\mathbf{Q}^{\mathrm{H}}\mathbf{a}_{\mathrm{r}}\left( \varTheta _{k}^{\mathrm{r}},\varPhi _{k}^{\mathrm{r}} \right) \mathbf{a}_{\mathrm{t}}\left( \varTheta _{k}^{\mathrm{t}},\varPhi _{k}^{\mathrm{t}} \right) ^H\mathbf{F1}\\
				\times e^{-j2\pi m\Delta f\left( \tau _k+{\delta}^{\mathrm{t}} \right)}e^{j2\pi \left( f_{k}^{\mathrm{d}}+{\delta}^{\mathrm{f}} \right) nT_{\mathrm{s}}}\\
			\end{array} \right]}+\hat{\mathbf{n}}_{m,n}.
	\end{align}
	By stacking the $N$ OFDM symbols and $M$ subcarriers, the received signal can be transformed to a three-order tensor $\mathcalbf{Y} \in \mathbb{C}^{R \times N \times M}$ as follows \cite{GaoffTensor}
	\begin{equation}\label{equation:tensor_Y_by_vector}
		\mathcalbf{Y} =\sum_{k=1}^{K+1}{\alpha _k\mathbf{a}(\varTheta _{k}^{\mathrm{r}},\varPhi _{k}^{\mathrm{r}},\varTheta _{k}^{\mathrm{t}},\varPhi _{k}^{\mathrm{t}})\circ \mathbf{b}(f_{k}^{\mathrm{d}},{\delta}^{\mathrm{f}})\circ \mathbf{c}(\tau _k,{\delta}^{\mathrm{t}})}+\mathcalbf{N} ,
	\end{equation}
	where $\mathcalbf{N}$ is the noise tensor, $\mathbf{a}\left( \varTheta _{k}^{\mathrm{r}},\varPhi _{k}^{\mathrm{r}},\varTheta _{k}^{\mathrm{t}},\varPhi _{k}^{\mathrm{t}} \right)  \in \mathbb{C}^{R  \times 1}$, $\mathbf{b}\left( f_{k}^{\mathrm{d}},{\delta}^{\mathrm{f}} \right) \in \mathbb{C}^{N \times 1}$, and $\mathbf{c}\left( \tau _k,{\delta}^{\mathrm{t}} \right)  \in \mathbb{C}^{M \times 1}$ can be respectively derived as follows
	\begin{align}
		\!\mathbf{a}\left( \varTheta _{k}^{\mathrm{r}},\varPhi _{k}^{\mathrm{r}},\varTheta _{k}^{\mathrm{t}},\varPhi _{k}^{\mathrm{t}} \right) &=\mathbf{Q}^{\mathrm{H}}\mathbf{a}_r\left( \varTheta _{k}^{\mathrm{r}},\varPhi _{k}^{\mathrm{r}} \right) \mathbf{a} ^{\mathrm{H}}_t\left( \varTheta _{k}^{\mathrm{t}},\varPhi _{k}^{\mathrm{t}} \right)\mathbf{F1}, \!\label{vector_a}
		\\
		\mathbf{b}(f_{k}^{\mathrm{d}},{\delta}^{\mathrm{f}})&=[1,\cdots ,e^{j2\pi (N-1)T_{\mathrm{s}}(f_{k}^{\mathrm{d}}+{\delta}^{\mathrm{f}})}]^{\mathrm{T}},\label{vector_b}
		\\
		\mathbf{c}\left( \tau _k,{\delta}^{\mathrm{t}} \right) &=[1,\cdots ,e^{-j2\pi (M-1)\Delta f(\tau _k+{\delta}^{\mathrm{t}})}]^{\mathrm{T}}.\label{vector_c}
	\end{align}
	By defining factor matrices $\mathbf{A}\in \mathbb{C}^{R \times (K+1)}$, $\mathbf{B}\in \mathbb{C}^{N \times (K+1)}$, and $\mathbf{C}\in \mathbb{C}^{M \times (K+1)}$, whose $k$th column are respectively given by $\mathbf{a}\left( \varTheta _{k}^{\mathrm{r}},\varPhi _{k}^{\mathrm{r}},\varTheta _{k}^{\mathrm{t}},\varPhi _{k}^{\mathrm{t}} \right)$, $\alpha _k \mathbf{b}(f_{k}^{\mathrm{d}},{\delta}^{\mathrm{f}})$, and $\mathbf{c}\left( \tau _k,{\delta}^{\mathrm{t}} \right)$, \eqref{equation:tensor_Y_by_vector} can also be expressed as 
	\begin{align}\label{equation:tensor_Y_by_matrix}
		\mathcalbf{Y} =\llbracket \mathbf{A},\mathbf{B},\mathbf{C} \rrbracket+\mathcalbf{N}.
	\end{align}
	
	\subsection{Decomposition-Based Factor Matrices Recovery}\label{subsec_Matrices_Recovery}
	From \eqref{vector_a}-\eqref{vector_c}, it can be observed that the factor matrices $\mathbf{A}$, $\mathbf{B}$, and $\mathbf{C}$ respectively encode the angle, Doppler, and delay parameters of all propagation paths.
	To extract these sensing parameters, we first formulate the following CP decomposition problem to recover $\mathbf{A}$, $\mathbf{B}$, and $\mathbf{C}$:
	\begin{equation}\label{Proble_CP_decomp}
		\min_{\mathbf{A},\mathbf{B},\mathbf{C}} \left\| \mathcalbf{Y} -\sum_{k=1}^{K+1}{\alpha _k\mathbf{a}(\varTheta _{k}^{\mathrm{r}},\varPhi _{k}^{\mathrm{r}},\varTheta _{k}^{\mathrm{t}},\varPhi _{k}^{\mathrm{t}})\!\circ\! \mathbf{b}(f_{k}^{\mathrm{d}},{\delta}^{\mathrm{f}})\! \circ \! \mathbf{c}(\tau _k,{\delta}^{\mathrm{t}})} \right\| _{\mathrm{F}}^{2},
	\end{equation}
	where $\|\cdot\|_{\mathrm{F}}$ denotes the Frobenius norm. 
	While conventional alternating least squares (ALS) methods \cite{zhouzhou,C-DRCNN} can solve the non-convex Problem \eqref{Proble_CP_decomp}, they suffer from computational inefficiency, initialization-dependent convergence, and numerical instability.
	To address these limitations, in the following, we leverage the inherent Vandermonde structure of delay-related matrix $\mathbf{C}$ through an ESPRIT-inspired decomposition scheme. 
	
	Let $\mathbf{Y}_{(1)}$ represent the mode-1 unfolding of tensor $\mathcalbf{Y}$, expressed as
	\begin{equation}\label{equation:Y1}
		\mathbf{Y}_{(1)}^{\mathrm{T}}=\left(\mathbf{C} \odot \mathbf{B}\right)\mathbf{A}^{\mathrm{T}} \in \mathbb{C}^{M N \times R}.
	\end{equation}
	In addition, we select a pair of integer ${L_1,L_2}$ subjected to $L_1+L_2 = M+1$ and define the cyclic matrix as 
	\begin{equation}
		\mathbf{J}_{l}=\left[ \mathbf{0}_{L_1\times (l-1)},\mathbf{I}_{L_1},\mathbf{0}_{L_1\times (L_2-l)} \right] \in \mathbb{C} ^{L_1\times M}.
	\end{equation}
	Then, a smoothed measurement matrix can be constructed as 
	\begin{align}\label{equation:Y_S}
		\mathbf{Y}_{\mathrm{S}}&=\left[ \left( \mathbf{J}_1\otimes \mathbf{I}_N \right) \mathbf{Y}_{\left( 1 \right)}^{\mathrm{T}},\dots ,\left( \mathbf{J}_{L_2}\otimes \mathbf{I}_N \right) \mathbf{Y}_{\left( 1 \right)}^{\mathrm{T}} \right]\notag
		\\
		&\overset{\left( a \right)}{=}\left( \mathbf{C}^{L_1}\odot \mathbf{B} \right) \left( \mathbf{C}^{L_2}\odot \mathbf{A} \right) ^{\mathrm{T}}\in \mathbb{C} ^{L_1N\times L_2R},
	\end{align}
	where $\mathbf{C}^{L_1}= [\mathbf{C}]_{1:L_1,:}$ and $\mathbf{C}^{L_2}=[\mathbf{C}]_{1:L_2,:}$ denote the first $L_1$ and $L_2$ rows of $\mathbf{C}$, respectively. 
	The equality $(a)$ follows from Khatri-Rao product properties, i.e., $(\mathbf{A} \otimes \mathbf{B})(\mathbf{C} \odot \mathbf{D})=(\mathbf{A}\mathbf{C}) \odot (\mathbf{B}\mathbf{D})$, and the shift-invariance induced by the Vandermonde structure of $\mathbf{C}$, i.e., $\mathbf{J}_{l_2} \mathbf{C}=\mathbf{J}_1 \mathbf{C} \mathrm{diag}\left(\left[\mathbf{C}\right]_{l_2,:}\right)$ \cite{timevarying_TVT}.
	By performing truncated singular value decomposition (SVD) on $\mathbf{Y}_{\mathrm{S}}$, we have
	\begin{equation}
		\mathbf{Y}_{\mathrm{S}}=\mathbf{U} \mathbf{\Sigma} \mathbf{V}^{\mathrm{H}},
		\label{equation:svd_Y_S}
	\end{equation}
	where $\mathbf{U}\in \mathbb{C}^{L_1N \times (K+1)}$, $\mathbf{\Sigma} \in \mathbb{C}^{(K+1) \times (K+1)}$, and $\mathbf{V} \in \mathbb{C}^{L_2R \times (K+1)}$. 
	Note that $\mathbf{U}$ spans the column spaces of  $\mathbf{Y}_S$. 
	Therefore, a nonsingular matrix $\mathbf{M} \in \mathbb{C}^{(K+1) \times (K+1)}$ exists such that \cite{zxd}
	\begin{align}
		\mathbf{C}^{L_1} \odot \mathbf{B}&=\mathbf{U M} \in \mathbb{C}^{L_1 N \times (K+1)},	\label{equation:UM}
		\\
		\mathbf{C}^{L_2} \odot \mathbf{A}&=\mathbf{V}^*\mathbf{\Sigma }\left( \mathbf{M}^{-1} \right) ^{\mathrm{T}}\in \mathbb{C} ^{L_2R\times (K+1)}. \label{equation:VSigmaM}
	\end{align}
	By leveraging the shift-invariant structure, we partition $\mathbf{U}$ into overlapped submatrices $\mathbf{U}_1={\left[ \mathbf{U}\right] }_{1:(L_1-1)N,:}$ and  $ \mathbf{U}_2=\left[ \mathbf{U}\right] _{N+1:L_1N,:}$, yielding
	\begin{align}
		\mathbf{U}_1 \mathbf{M}& = \underline{\mathbf{C}}^{L_1} \odot \mathbf{B}, 
		\label{equation:up}
		\\
		\mathbf{U}_2 \mathbf{M}& = \overline{\mathbf{C}}^{L_1} \odot \mathbf{B},
		\label{equation:down}
	\end{align}
	where $\underline{\mathbf{C}}^{L_1}$ and $\overline{\mathbf{C}}^{L_1}$ respectively denote $\mathbf{C}^{L_1}$ with the first and last rows removed.  
	Based on the Vandermonde structure of $\mathbf{C}^{L_1}$, we derive that 
	\begin{equation}
		(\underline{\mathbf{C}}^{L_1} \odot \mathbf{B})\mathbf{Z}= \overline{\mathbf{C}}^{L_1} \odot \mathbf{B},
		\label{equation:Vandermonde}
	\end{equation}
	where $\mathbf{Z}= \mathrm{diag}([e^{-j 2 \pi \Delta f \Delta \tau_1}, \ldots, e^{-j 2 \pi \Delta f \Delta \tau_K}])$.
	Substituting \eqref{equation:up} and \eqref{equation:down} into \eqref{equation:Vandermonde}, we have
	\begin{equation}
		\mathbf{U}_1^{\dagger} \mathbf{U}_2 =\mathbf{M Z M}^{-1}.
		\label{equation:performing_svd}
	\end{equation}

	From \eqref{equation:performing_svd}, it can be noted  that the estimates of $\mathbf{Z}$ and $\mathbf{M}$, denoted by $\hat{\mathbf{Z}}$ and $\hat{\mathbf{M}}$, can be obtained from the eigenvalue decomposition (EVD) of $\mathbf{U}_1^{\dagger}\mathbf{U}_2$. 
	Then, a recovery of $\mathbf{C}$ can be generated by the normalized diagonal elements of $\hat{\mathbf{Z}}$.
	Specifically, each column of the recovered factor matrix ${\mathbf{C}}$ is given by
	\begin{equation}
		\hat{\mathbf{c}}_k =\left[1, \hat{z}_k, \ldots, \hat{z}_k^{M-1}\right]^{\mathrm{T}},\,k=1, \ldots, K+1,
		\label{equation:recovered_C}
	\end{equation}
	where $\hat{z}_k={[\hat{\mathbf{Z}}]_{k,k}}\big/{\left| [\hat{\mathbf{Z}}]_{k,k} \right|}$.
	Recall that the $k$th column of $\mathbf{B}$ is given by $\alpha _k \mathbf{b}(f_{k}^{\mathrm{d}},{\delta}^{\mathrm{f}})$.
	Let $\mathbf{c}_{k}^{L_1}$ and $\mathbf{m}_k$ denote the $k$th columns of $\mathbf{C}^{L_1}$ and $\mathbf{M}_k$, respectively.
	From the Kronecker product structure in \eqref{equation:UM}, we have 
	\begin{equation}
		\mathbf{c}_{k}^{L_1}\otimes \left( \alpha _k\mathbf{b}(f_{k}^{\mathrm{d}},{\delta}^{\mathrm{f}}) \right) =\mathbf{Um}_k.
	\end{equation}
	Hence, we can further recover each column of $\mathbf{B}$ as follows
	\begin{align}
		\hat{\mathbf{b}}_k\overset{\left( a \right)}{=} &\left( \frac{(\hat{\mathbf{c}}_{k}^{L_1})^{\mathrm{H}}}{\left\| \hat{\mathbf{c}}_{k}^{L_1} \right\| _{2}^{2}}\otimes \mathbf{I}_N \right) \left( \hat{\mathbf{c}}_{k}^{L_1}\otimes \hat{\mathbf{b}}_k \right) \notag
		\\
		=&\left( \frac{(\hat{\mathbf{c}}_{k}^{L_1})^{\mathrm{H}}}{\left\| \hat{\mathbf{c}}_{k}^{L_1} \right\| _{2}^{2}}\otimes \mathbf{I}_N \right) \mathbf{U}\hat{\mathbf{m}}_k,\,k=1,\cdots ,K+1,
		\label{equation:recovered_B}
	\end{align} 
	where $\hat{\mathbf{c}}_{k}^{L_1}$ and $\hat{\mathbf{m}}_k$ represent the $k$th columns of $\hat{\mathbf{C}}^{L_1}$ and $\hat{\mathbf{M}}_k$, respectively.
	The equality $\left( a \right)$ in \eqref{equation:recovered_B} follows from the Kronecker product identities, i.e., $\left( \mathbf{A}\otimes \mathbf{B} \right) \left( \mathbf{C}\otimes \mathbf{D} \right) =\mathbf{AC}\otimes \mathbf{BD}$.
	Similarly, it can be derived from \eqref{equation:UM} that $\mathbf{c}_k^{L_2} \otimes \mathbf{a}_k = \mathbf{V^*}\mathbf{\Sigma}\mathbf{t}_k$, where $\mathbf{t}_k$ is the $k$th column of $\left( \mathbf{M}^{-1} \right) ^{\mathrm{T}}$.
	Thus, each column of ${\mathbf{A}}$ can be recovered as 
	\begin{equation}
		\hat{\mathbf{a}}_k=\left(\frac{(\hat{\mathbf{c}}_k^{L_2})^{\mathrm{H}}}{\left\|\hat{\mathbf{c}}_k^{L_2}\right\|_2^2} \otimes \mathbf{I}_R\right) \mathbf{V}^{*} \mathbf{\Sigma} \hat{\mathbf{t}}_k, \,k = 1,\cdots,K+1,
		\label{equation:recovered_A}
	\end{equation} 
	where $\hat{\mathbf{c}}_{k}^{L_2}$ and $\hat{\mathbf{t}}_k$ represent the $k$th columns of $\hat{\mathbf{C}}^{L_2}$ and $( \hat{\mathbf{M}}^{-1} ) ^{\mathrm{T}}$, respectively.
	
	\begin{remark}
		Here, the factor matrices $\mathbf{A}$, $\mathbf{B}$, and $\mathbf{C}$ are recovered as $\hat{\mathbf{A}}=\left[ \hat{\mathbf{a}}_1,\dots ,\hat{\mathbf{a}}_{K+1} \right] $, $\hat{\mathbf{B}}=\left[ \hat{\mathbf{b}}_1,\dots ,\hat{\mathbf{b}}_{K+1} \right] $, and $\hat{\mathbf{C}}=\left[ \hat{\mathbf{c}}_1,\dots ,\hat{\mathbf{c}}_{K+1} \right] $, respectively.
		The uniqueness of the CP decomposition will be discussed in Section \ref{subsec_uniqueness}, which	
		guarantees the following equalities \cite{AlwinOnKruskal}:
		\begin{equation}\label{equation_uniqueness}
			\hat{\mathbf{A}}=\mathbf{A} \mathbf{\Pi} \mathbf{\Delta}_1, \
			\hat{\mathbf{B}}=\mathbf{B} \mathbf{\Pi} \mathbf{\Delta}_2, \  \hat{\mathbf{C}}=\mathbf{C} \mathbf{\Pi} \mathbf{\Delta}_3,
		\end{equation}
		where $\mathbf{\Pi}$ is a unique permutation matrix, while $\mathbf{\Delta}_1$, $\mathbf{\Delta}_2$, and $\mathbf{\Delta}_3$ are unique diagonal scaling matrices satisfying $\mathbf{\Delta }_1\mathbf{\Delta }_2\mathbf{\Delta }_3=\mathbf{I}_{K+1}$. 
		The shared permutation matrix ensures automatic parameter pairing - the $k$th columns across all factor matrices inherently correspond to the same propagation path, which reduces the data association requirements.
	\end{remark}

	\subsection{Parameter Estimation}\label{subsec_Parameter_Estimation}
	In this subsection, we extract the bistatic ranges, azimuth/elevation angles, and bistatic Doppler velocities correspond to all targets from the recovered factor matrices $\hat{\mathbf{A}}$, $\hat{\mathbf{B}}$, and $\hat{\mathbf{C}}$.
	
	First, the delay of the $k$th path can be estimated as 
	\begin{equation}
		\Delta \hat{\tau}_k\triangleq\hat{\tau}_k+\hat{\delta}^{\mathrm{t}}=\frac{\measuredangle \hat{z}_k}{-2\pi \Delta f}, \,k = 1,\cdots,K+1,
	\end{equation}
	where $\hat{\delta}^{\mathrm{t}}$ denotes the estimate of the STO ${\delta}^{\mathrm{t}}$, and $\measuredangle \hat{z}_k$ is the angle of $\hat{z}_k$. 
	Without estimating the permutation matrix $\mathbf{\Pi}$, we can index the minimum value of $\{\Delta \hat{\tau}_k\}_{k=1}^{K+1}$ to find the LoS path, i.e.,
	\begin{equation}
		k_{\mathrm{LoS}} = \arg\min_{\substack{1 \leqslant k \leqslant K+1}} \Delta \hat{\tau}_k.
		\label{equation:LoS_path}
	\end{equation}
	Since the locations of all BSs are apriori known, the residual STO can be estimated as 
	\begin{equation} 
		\hat{\delta}^t = \Delta \hat{\tau}_{k_{\mathrm{LoS}}}-{\tau}_{k_{\mathrm{LoS}}},
	\end{equation} 
	where $\tau_{k_{\mathrm{LoS}}}= d_{k_{\mathrm{LoS}}}/c_0$, and $d_{k_{\mathrm{LoS}}}$ is the distance between tBS $n_{\mathrm{t}}$ and rBS $n_{\mathrm{r}}$.
	Then, the delay of the target-scattered paths can be extracted as 
	\begin{equation} 
		\hat{\tau}_k = \Delta \hat{\tau}_k-\hat{\delta}^t,\, k \in \{1, \cdots, K+1\}\setminus k_{\mathrm{LoS}}.
	\end{equation}
	The corresponding bistatic ranges are obtained as 
	\begin{equation}
		\hat{d}_k=\hat{\tau}_k c_0, \, k \in \{1, \cdots, K+1\}\setminus k_{\mathrm{LoS}}.
		\label{equation:estimated_bistatic_range}
	\end{equation}
	
	Next, the estimation of the Doppler parameters $\{f^{\mathrm{d}}_k,{\delta}^{\mathrm f}\}$ can be deduced as 
	\begin{align}
		&\Delta \hat{f}_{k}^{\mathrm{d}}\triangleq \hat{\delta}^{\mathrm{f}}+\hat{f}_{k}^{\mathrm{d}}=\mathrm{arg}\max_{\Delta f_{k}^{\mathrm{d}}} \frac{\left| \hat{\mathbf{b}}_{k}^{\mathrm{H}}\mathbf{b}\left( f_{k}^{\mathrm{d}},\delta ^{\mathrm{f}} \right) \right|^2}{\left\| \hat{\mathbf{b}}_k \right\| _{2}^{2}\left\| \mathbf{b}\left( f_{k}^{\mathrm{d}},\delta ^{\mathrm{f}} \right) \right\| _{2}^{2}},\notag
		\\
		&k \in \{1, \cdots, K+1\}\setminus k_{\mathrm{LoS}}.
		\label{equation:estimate_sumDroppler}
	\end{align}
	The solution of Problem \eqref{equation:estimate_sumDroppler} can be obtained via the one-dimensional search.
	By utilizing $f_{k_{\mathrm{LoS}}}^{\mathrm{d}} =0$, we can separate $\hat{f}_{k}^{\mathrm{d}}$ and $\hat{\delta}^{\mathrm f}$ as follows
	\begin{align} 
		&\hat{\delta}^{\mathrm{f}}=\Delta \hat{f}_{k_{\mathrm{LoS}}}^{\mathrm{d}},
		\\
		&\hat{f}_{k}^{\mathrm{d}}=\Delta \hat{f}_{k}^{\mathrm{d}}-\hat{\delta}^{\mathrm{f}},\,k\in \{1,\cdots ,K+1\}\setminus k_{\mathrm{LoS}}.
	\end{align}
	Then, the bistatic Doppler velocities are estimated as 
	\begin{equation}
		\hat{v}_k=\hat{f}^{\mathrm{d}}_k \lambda,\, k \in \{1, \cdots, K+1\}\setminus k_{\mathrm{LoS}}.
		\label{equation:estimated_bistatic_Doppler_velocity}
	\end{equation}
	
	Finally, we extract the angle parameters from $\hat{\mathbf{A}}$.
	Note that it is challenging for the four-dimensional parameter extraction of $\left\{ \theta _{k}^{\mathrm{r}},\phi _{k}^{\mathrm{r}},\theta _{k}^{\mathrm{t}},\phi _{k}^{\mathrm{t}} \right\} _{k=1}^{K+1}$.
	Therefore, we propose to only estimate the AoAs $\{\theta _{k}^{\mathrm{r}},\phi _{k}^{\mathrm{r}}\}_{k=1}^{K+1}$.
	The AoDs $\{\theta _{k}^{\mathrm{t}},\phi _{k}^{\mathrm{t}}\}_{k=1}^{K+1}$ can be calculated after target localization.
	From \eqref{vector_a}, we formulate the problem for jointly estimating  ${\varTheta}_{k}^{\mathrm{r}}$ and ${\varPhi}_{k}^{\mathrm{r}}$ as
	\begin{equation}
		\max_{\varTheta _{k}^{\mathrm{r}},\varPhi _{k}^{\mathrm{r}}} \frac{\left| \hat{\mathbf{a}}_{k}^{\mathrm{H}}\mathbf{Q}^{\mathrm{H}}\mathbf{a}_{\mathrm{r}}\left( \varTheta _{k}^{\mathrm{r}},\varPhi _{k}^{\mathrm{r}} \right) \right|^2}{\left\| \hat{\mathbf{a}}_k \right\| _{2}^{2}\left\| \mathbf{Q}^{\mathrm{H}}\mathbf{a}_{\mathrm{r}}\left( \varTheta _{k}^{\mathrm{r}},\varPhi _{k}^{\mathrm{r}} \right) \right\| _{2}^{2}},
		\label{Prob_2D_AOA}
	\end{equation}
	where the normalization operation eliminates the scaling ambiguity caused by $\mathbf{\Delta}_1$ and the impact of AoDs. 
	In the objective function of Problem \eqref{Prob_2D_AOA}, the variables ${\varTheta}_{k}^{\mathrm{r}}$ and ${\varPhi}_{k}^{\mathrm{r}}$ are highly coupled.
	Although the problem can be solved via a two-dimensional search, the complexity is prohibitively high.
	In the following, an SVD-based decoupling algorithm is proposed for estimating $\{\theta _{k}^{\mathrm{r}},\phi _{k}^{\mathrm{r}}\}_{k=1}^{K+1}$ efficiently.
	First, we approximate the combining matrix $\mathbf{Q}$ as the Kronecker product of two matrices by solving the following problem \cite{VanLoan1993}:
	\begin{equation}\label{Problem_aproxQ}
		\min_{\mathbf{Q}_{\mathrm{v}},\mathbf{Q}_{\mathrm{h}}} \parallel \mathbf{Q}-\mathbf{Q}_{\mathrm{v}}\otimes \mathbf{Q}_{\mathrm{h}}\parallel _{\mathrm{F}}^{2},
	\end{equation}
	where $\mathbf{Q}_{\mathrm{v}} \in \mathbb{C}^{N_{\mathrm{v}} \times J_{\mathrm{v}}}$, $\mathbf{Q}_{\mathrm{h}} \in \mathbb{C}^{N_{\mathrm{h}} \times J_{\mathrm{h}}}$, and $J_{\mathrm{v}} J_{\mathrm{h}} = R$. 
	Problem \eqref{Problem_aproxQ} can be equivalently reformulated as 
	\begin{equation}\label{Problem_aproxQ_reform}
		\min_{\mathbf{Q}_{\mathrm{v}},\mathbf{Q}_{\mathrm{h}}} \left\| R\left( \mathbf{Q} \right) -\mathrm{vec}\left( \mathbf{Q}_{\mathrm{v}} \right)  \mathrm{vec}\left( \mathbf{Q}_{\mathrm{h}} \right) ^{\mathrm{T}} \right\| _{\mathrm{F}}^{2},
	\end{equation}
	where $R\left( \mathbf{Q} \right)$ is defined as follows:
	\begin{equation} \label{R_Q}
		\!\! R\left( \mathbf{Q} \right) =\left[ \begin{array}{c}
			\mathbf{Q}_1\\
			\vdots\\
			\mathbf{Q}_{j_{\mathrm{v}}}\\
			\vdots\\
			\mathbf{Q}_{J_{\mathrm{v}}}\\
		\end{array} \right] \!,\ 
		\mathbf{Q}_{j_{\mathrm{v}}}=\left[ \begin{array}{c}
			\mathrm{vec}\left( \left[ \mathbf{Q} \right] _{1,j_{\mathrm{v}}}\mathbf{Q}_{\mathrm{h}} \right) ^{\mathrm{T}}\\
			\vdots\\
			\mathrm{vec}\left( \left[ \mathbf{Q} \right] _{N_{\mathrm{v}},j_{\mathrm{v}}}\mathbf{Q}_{\mathrm{h}} \right) ^{\mathrm{T}}\\
		\end{array} \right] \!.
	\end{equation}
	The rank-one approximation problem \eqref{Problem_aproxQ_reform} can be solved by performing SVD on $R\left( \mathbf{Q} \right)$ \cite{eckart1936approximation}.
	Let $\sigma _{\max}$ denote the maximum singular value of $R\left( \mathbf{Q} \right)$.
	Then, the optimal solutions of Problem \eqref{Problem_aproxQ_reform} are given by
	\begin{align}
		\mathbf{Q}_{\mathrm{v}}&=\mathrm{unvec}_{N_{\mathrm{v}}\times J_{\mathrm{v}}}\left( \sqrt{\sigma _{\max}}\mathbf{u}_{\max} \right) ,
		\label{Q_v}
		\\
		\mathbf{Q}_{\mathrm{h}}&=\mathrm{unvec}_{N_{\mathrm{h}}\times J_{\mathrm{h}}}\left( \sqrt{\sigma _{\max}}\mathbf{v}_{\max}^{*} \right) ,
		\label{Q_h}
	\end{align}
	where ${\mathbf{u}}_{\max}$ and ${\mathbf{v}}_{\max}$ represent the left and right singular vectors corresponding to $\sigma _{\max}$, respectively.
	Thus, \eqref{vector_a} can be rewritten as follows
	\begin{align}
		&\quad \mathbf{a}\left( \varTheta _{k}^{\mathrm{r}},\varPhi _{k}^{\mathrm{r}},\varTheta _{k}^{\mathrm{t}},\varPhi _{k}^{\mathrm{t}} \right) \notag 
		\\
		&=\left( \mathbf{Q}_{v}^{\mathrm{H}}\otimes \mathbf{Q}_{h}^{\mathrm{H}} \right) \left( \mathbf{a}_{r}^{\mathrm{v}}\left( \varPhi _{k}^{\mathrm{r}} \right) \otimes \mathbf{a}_{r}^{\mathrm{h}}\left( \varTheta _{k}^{\mathrm{r}},\varPhi _{k}^{\mathrm{r}} \right) \right) \mathbf{a}_{t}^{\mathrm{H}}\left( \varTheta _{k}^{\mathrm{t}},\varPhi _{k}^{\mathrm{t}} \right) \mathbf{F1}\notag
		\\
		&=\left( \mathbf{Q}_{v}^{\mathrm{H}}\mathbf{a}_{r}^{\mathrm{v}}\left( \varPhi _{k}^{\mathrm{r}} \right) \right) \otimes \left( \mathbf{Q}_{h}^{\mathrm{H}}\mathbf{a}_{r}^{\mathrm{h}}\left( \varTheta _{k}^{\mathrm{r}},\varPhi _{k}^{\mathrm{r}} \right) \right) \mathbf{a}_{t}^{\mathrm{H}}\left( \varTheta _{k}^{\mathrm{t}},\varPhi _{k}^{\mathrm{t}} \right) \mathbf{F1}\notag
		\\
		&\triangleq \mathbf{o}_k\otimes \mathbf{w}_k,
	\end{align}
	where 
	\vspace{-0.2cm}
	\begin{align}
		&\mathbf{o}_k=\mathbf{Q}_{v}^{\mathrm{H}}\mathbf{a}_{r}^{\mathrm{v}}\left( \varPhi _{k}^{\mathrm{r}} \right),
		\label{equation:ok}\\
		&\mathbf{w}_k=\mathbf{Q}_{h}^{\mathrm{H}}\mathbf{a}_{r}^{\mathrm{h}}\left( \varTheta _{k}^{\mathrm{r}},\varPhi _{k}^{\mathrm{r}} \right) \mathbf{a}_{t}^{\mathrm{H}}\left( \varTheta _{k}^{\mathrm{t}},\varPhi _{k}^{\mathrm{t}} \right) \mathbf{F1}. 
		\label{equation:wk}
	\end{align}
	Moreover, by defining $\mathbf{O}=\left[ \mathbf{o}_1,\dots ,\mathbf{o}_{K+1} \right] $ and $\mathbf{W}=\left[ \mathbf{w}_1,\dots ,\mathbf{w}_{K+1} \right] $, we can reformulate factor matrix $\mathbf{A}$ as follows
	\vspace{-0.2cm}
	\begin{align}
		\mathbf{A}&=[\mathbf{o}_1\otimes \mathbf{w}_1,\cdots ,\mathbf{o}_{K+1}\otimes \mathbf{w}_{K+1}] \notag
		\\
		&= \mathbf{O}\odot \mathbf{W}.
	\end{align}
	To estimate $\mathbf{O}=\left[ \mathbf{o}_1,\dots ,\mathbf{o}_{K+1} \right] $ and $\mathbf{W}=\left[ \mathbf{w}_1,\dots ,\mathbf{w}_{K+1} \right] $, we formulate the following problem
	\begin{equation}
		\max_{\mathbf{o}_k,\mathbf{w}_k} \left\| \mathrm{unvec}_{J_{\mathrm{v}}\times J_{\mathrm{h}}}(\hat{\mathbf{a}}_k)-\mathbf{o}_k\mathbf{w}_{k}^{\mathrm{T}} \right\| _{\mathrm{F}}^{2}.
		\label{equation:SVDone}
	\end{equation}
	Problem \eqref{equation:SVDone} can be efficiently solved via the rank-1 truncated SVD.
	Let $\sigma _{k,\max}$, $\mathbf{u}_{k,\max}$, and $\mathbf{v}_{k,\max}$ denote the maximum singular value, and the corresponding left and right singular vectors, respectively, of $\mathrm{unvec}_{J_{\mathrm{v}}\times J_{\mathrm{h}}}(\hat{\mathbf{a}}_k)$.
	We can set the solution of Problem \eqref{equation:SVDone} as
	\begin{equation}\label{hat_ow}
		\hat{\mathbf{o}}_k=\sigma _{k,\max}\mathbf{u}_{k,\max},
		\
		\hat{\mathbf{w}}_k=\mathbf{v}^*_{k,\max}.
	\end{equation}
	Then, Problem \eqref{Prob_2D_AOA} can be decoupled as two subproblems:
	\begin{equation}
		\max_{\varPhi _{k}^{\mathrm{r}}} \frac{\left| \hat{\mathbf{o}}_{k}^{\mathrm{H}}\left( \mathbf{Q}_{\mathrm{v}}^{\mathrm{H}}\mathbf{a}_{r}^{\mathrm{v}}\left( \varPhi _{k}^{\mathrm{r}} \right) \right) \right|^2}{\left\| \hat{\mathbf{o}}_k \right\| _{2}^{2}\left\| \left( \mathbf{Q}_{\mathrm{v}}^{\mathrm{H}}\mathbf{a}_{r}^{\mathrm{v}}\left( \varPhi _{k}^{\mathrm{r}} \right) \right) \right\| _{2}^{2}},
		\label{equation:one_search_Phi}
	\end{equation}
	\begin{equation}
		\max_{\varTheta _{k}^{\mathrm{r}}} \frac{\left| \hat{\mathbf{w}}_{k}^{\mathrm{H}}\left( \mathbf{Q}_{\mathrm{h}}^{\mathrm{H}}\mathbf{a}_{r}^{\mathrm{h}}\left( \varTheta _{k}^{\mathrm{r}},\hat{\varPhi}_{k}^{\mathrm{r}} \right) \right) \right|^2}{\left\| \hat{\mathbf{w}}_k \right\| _{2}^{2}\left\| \left( \mathbf{Q}_{\mathrm{h}}^{\mathrm{H}}\mathbf{a}_{r}^{\mathrm{h}}\left( \varTheta _{k}^{\mathrm{r}},\hat{\varPhi}_{k}^{\mathrm{r}} \right) \right) \right\| _{2}^{2}}.
		\label{equation:one_search_Theta}
	\end{equation}
	These subproblems can be solved sequentially through one-dimensional search method.
	From \eqref{equation:azimuth/elevationAoA}, given the optimal solutions $\hat{\varTheta}_{k}^{\mathrm{r}}$ and $\hat{\varTheta}_{k}^{\mathrm{r}}$ of subproblems \eqref{equation:one_search_Phi} and \eqref{equation:one_search_Theta}, the azimuth and elevation angles of AoA can be calculated as
	\vspace{-0.1cm}
	\begin{equation}    
		\hat{\theta}_{k}^{\mathrm{r}}=\mathrm{arc}\cos ( \hat{\varPhi}_{k}^{\mathrm{r}} ) ,\mathrm{}\hat{\phi}_{k}^{\mathrm{r}}=\mathrm{arc}\cos \left(  \frac{\hat{\varTheta}_{k}^{\mathrm{r}}}{\sin ( \hat{\theta}_{k}^{\mathrm{r}} )} \right)  .
		\label{equation:azimuth_doa}
	\end{equation}
	In Algorithm \ref{alg_SVD_Decoup}, we summarize the procedures of the proposed SVD-based algorithm for solving Problem \eqref{Prob_2D_AOA} and estimating the AoAs $\{\theta _{k}^{\mathrm{r}},\phi _{k}^{\mathrm{r}}\}_{k=1}^{K+1}$.
	
	\begin{algorithm}[tb]
		\caption{SVD-based Decoupling AoA Estimation}
		\label{alg_SVD_Decoup}
		\begin{algorithmic}[1] 
			\Require 
			Combining matrix $\mathbf{Q}$ and recovered factor matrix $\hat{\mathbf{A}}$.
			\Ensure 
			The estimation of azimuth/elevation AoAs $\{\theta _{k}^{\mathrm{r}},\phi _{k}^{\mathrm{r}}\}_{k=1}^{K+1}$.
			\State \textbf{Initialization}: 
			Set $J_{\mathrm{v}}$ and $J_{\mathrm{h}}$ subject to $J_{\mathrm{v}} J_{\mathrm{h}} = R$;
			\State Calculate $R\left( \mathbf{Q} \right)$ via \eqref{R_Q}; 
			\State Obtain $\mathbf{Q}_{\mathrm{v}}$ and $\mathbf{Q}_{\mathrm{h}}$ via \eqref{Q_v} and \eqref{Q_h}, respectively; \label{step:q_design}
			\For{$k=1:K+1$} \label{step_loopbegin}
			\State Obtain $\hat{\mathbf{o}}_k$ and $\hat{\mathbf{w}}_k$ via \eqref{hat_ow}; \label{step:unvec}
			\State Obtain $\hat{\varPhi}_{k}^{\mathrm{r}}$ and $\hat{\varTheta}_{k}^{\mathrm{r}}$ by solving Problem \eqref{equation:one_search_Phi} and Problem \eqref{equation:one_search_Theta} via the one-dimensional search, respectively;
			\State Calculate $\hat{\theta}_{k}^{\mathrm{r}}$ and $\hat{\phi}_{k}^{\mathrm{r}}$ via \eqref{equation:azimuth_doa}; \label{step_cal_theta_phi}
			\EndFor \label{step_loopend}
		\end{algorithmic}
	\end{algorithm}

	\vspace{-0.2cm}
	\subsection{Uniqueness Analysis} \label{subsec_uniqueness}
	This subsection establishes uniqueness guarantees for the tensor decomposition \eqref{equation:tensor_Y_by_matrix} to ensure reliable recovery of factor matrices. 
	Consider a third-order tensor $\mathcal{X} \in \mathbb{C}^{I_1 \times I_2 \times I_3}$ with rank $K$ decomposed into factor matrices $\mathbf{A}^{(1)} \in \mathbb{C}^{I_1 \times K}$, $\mathbf{A}^{(2)} \in \mathbb{C}^{I_2 \times K}$ and $\mathbf{A}^{(3)} \in \mathbb{C}^{I_3 \times K}$, where $\mathbf{A}^{(3)}$ is a Vandermonde matrix with generators $\{z_k\}_{k=1}^K$.
	The uniqueness means that any possible combination of factor matrices $( \hat{\mathbf{A}}^{(1)},\mathrm{}\hat{\mathbf{A}}^{(2)},\hat{\mathbf{A}}^{(3)} ) $ for $\mathcal{X}$ satisfies
	\begin{equation}
		\!\!\!\hat{\mathbf{A}}^{\!(1)}=\mathbf{A}^{\!(1)}\mathbf{\Pi \Delta }_1 , \, \mathrm{}\hat{\mathbf{A}}^{\!(2)}=\mathbf{A}^{\!(2)}\mathbf{\Pi \Delta }_2, \, \hat{\mathbf{A}}^{\!(3)}=\mathbf{A}^{\!(3)}\mathbf{\Pi \Delta }_3,\! \!
	\end{equation}
	where $\mathbf{\Pi}$ is a $K\times K$ permutation matrix, and $\mathbf{\Delta}_1$, $\mathbf{\Delta}_2$, and $\mathbf{\Delta}_3$ with $\mathbf{\Delta }_1\mathbf{\Delta }_2\mathbf{\Delta }_3=\mathbf{I}_{K}$ are diagonal matrices. 
	By extending the classical Kruskal’s uniqueness theorem \cite{KRUSKAL197795} and incorporating structural constraints inherent to the factor matrices, the following relaxed uniqueness criterion has been proved in \cite[Theorem III.3]{blind_signal}:
	\begin{lemma}
		\emph{Let $k(\mathbf{A})$ denote the Kruskal-rank of matrix $\mathbf{A}$.
			The tensor decomposition of $\mathcal{X}$ is unique if the following joint rank conditions hold:
			\begin{equation}
				k(\mathbf{A}^{(L_1,3)} \odot \mathbf{A}^{(2)}) = K, 
			\end{equation}
			\begin{equation}
				k(\mathbf{A}^{(L_2,3)} \odot \mathbf{A}^{(1)}) = K,
			\end{equation}
			where $\mathbf{A}^{(L_1,3)}$ represents the first $L_1$ rows of $\mathbf{A}^{(3)}$, and $\mathbf{A}^{(L_2,3)}$ denotes the first $L_2$ rows of $\mathbf{A}^{(2)}$.
			These conditions are generically true if
			\begin{equation}
				\min\left(\left( L_1-1\right) I_2,\, L_2I_1\right) \geqslant K .
			\end{equation}
		}
		
	\end{lemma}

	In the considered system model, the tensor dimensions $I_1$, $I_2$ and $I_3$ correspond respectively to the number of RF chains $R$, OFDM symbols $N$ and subcarriers $M$.
	The rank of tensor $\mathcalbf{Y}$ is $K+1$.
	Therefore, sufficient conditions for the decomposition uniqueness can be formulated as follows
	\begin{align}
		L_1 + L_2 &= M + 1, \label{uniqueness_1}\\
		(L_1 - 1)N &\geqslant K+1, \label{uniqueness_2}\\
		L_2R &\geqslant K+1. \label{uniqueness_3}
	\end{align}
	These constraints can be satisfied by appropriately selecting parameters, ensuring automatic association of the extracted parameters with respective propagation paths.
	Note that telecommunication systems in 5G and beyond typically employ a large number of subcarriers $M$.
	This enables flexible selection of integers $L_1$ and $L_2$.
	
	\vspace{-0.2cm}
	\subsection{Algorithm Development}
	
	\begin{algorithm}[t]
		\caption{Tensor-Based Sensing Parameter Estimation}
		\label{alg_para_esti}
		\begin{algorithmic}[1]
			\Require 
			Received signal tensor $\mathcalbf{Y}$.
			\Ensure 
			Estimations of bistatic ranges, bistatic Doppler velocities, and azimuth/elevation angles of arrival $\left\{ d_k,v_k,\theta _{k}^{\mathrm{r}},\phi _{k}^{\mathrm{r}} \right\} _{k=1,k\ne k_{\mathrm{LoS}}}^{K+1}$.
			\State \textbf{Initialization}: 
			Mode-1 unfold tensor $\mathcalbf{Y}$ as $\mathbf{Y}_{(1)}$ in \eqref{equation:Y1}; Set smoothing parameters $ L_1$ and $L_2 $ subject to $L_1+L_2 = M+1$;
			\State Calculate the smoothed measurement matrix $\mathbf{Y}_{\mathrm{S}}$ via \eqref{equation:Y_S};
			\State Obtain $\mathbf{U}$, $\mathbf{\Sigma}$, and $\mathbf{V}$ by performing SVD on $\mathbf{Y}_{\mathrm{S}}$ in \eqref{equation:svd_Y_S};
			\State Obtain $\hat{\mathbf{Z}}$ and $\hat{\mathbf{M}}$ by performing EVD on $\mathbf{U}_1^{\dagger} \mathbf{U}_2$ in \eqref{equation:performing_svd};
			\State Construct columns of $\hat{\mathbf{C}}$ via \eqref{equation:recovered_C};
			\State Construct columns of $\hat{\mathbf{B}}$ via \eqref{equation:recovered_B};
			\State Construct columns of $\hat{\mathbf{A}}$ via \eqref{equation:recovered_A};
			\State Obtain the index of LoS path $k_{\mathrm{LoS}}$ via \eqref{equation:LoS_path}; \label{step_getkLoS}
			\For{$k \in \{1, \cdots, K+1\}\setminus k_{\mathrm{LoS}}$}
			\State Calculate $\hat{d}_k$ via \eqref{equation:estimated_bistatic_range};
			\State Calculate $\hat{v}_k$ via \eqref{equation:estimated_bistatic_Doppler_velocity};
			\State Obtain $\hat{\theta}_k$ and $\hat{\phi}_k$ following Step \ref{step:unvec}-\ref{step_cal_theta_phi} of Algorithm \ref{alg_SVD_Decoup};
			\EndFor
		\end{algorithmic}
	\end{algorithm}
	Based on the discussions in Sections \ref{subsec_Matrices_Recovery} and \ref{subsec_Parameter_Estimation}, we summarize the proposed tensor decomposition and spatial smoothing-based algorithm for sensing parameter estimation in Algorithm \ref{alg_para_esti}.
	
	\textit{Complexity Analysis:}
	The main computational complexity is analyzed as follows:
	\begin{itemize}
		\item \textbf{Step 2:} The computational complexity of constructing $\mathbf{Y}_{\mathrm{S}}$ is initially $\mathcal{O}(L_1L_2NR)$.
		However, by leveraging the sparse cyclic selection matrices $\mathbf{J}_{l}$ to extract and assemble specific rows from $\mathbf{Y}_{(1)}$ into the extended matrix formulation, this complexity can be reduced to practically negligible levels.
		
		\item \textbf{Step 3:} The complexity of performing truncated SVD on $\mathbf{Y}_S$ is $\mathcal{O}(L_1  L_2 N  R K )$.
		
		\item \textbf{Step 4:} Calculating the pseudo-inverse of $\mathbf{U}_1$ has a complexity of $\mathcal{O}(L_1 N K)$.
		Besides, performing EVD on $\mathbf{U}_1^{\dagger} \mathbf{U}_2$ has a complexity of $\mathcal{O}(K^3)$.
		
		\item \textbf{Step 5-7:} The construction of matrices $\mathbf{C}$, $\mathbf{B}$, and $\mathbf{A}$ primarily requires the complexity of $\mathcal{O}(K M)$, $\mathcal{O}(L_1N^2 K)$, and $\mathcal{O}(L_2R^2 K)$, respectively.
		
		\item \textbf{Step 9-13:} Denote by $G$ the number of iterations in the one-dimensional search method.
		In the parameter extraction stage, the complexity is dominated by that of estimating $\{\theta _{k}^{\mathrm{r}},\phi _{k}^{\mathrm{r}}\}_{k=1,k\ne k_{\mathrm{LoS}}}^{K+1}$, which is given by $\mathcal{O}(RKG(J_{\mathrm{v}}N_{\mathrm{v}}+J_{\mathrm{h}}N_{\mathrm{h}}))$.
	\end{itemize}
	In general, the total complexity of Algorithm \ref{alg_para_esti} can be summarized as $\mathcal{O} (L_1L_2NRK+L_1N^2K+L_2R^2K+RKG(J_{\mathrm{v}}N_{\mathrm{v}}+J_{\mathrm{h}}N_{\mathrm{h}}))$.
	
	\begin{remark}
		In the previous analysis, $k\in \left\{ 1,\dots ,K+1 \right\} $ has been used to index the propagation paths between tBS $n_{\mathrm{t}}$ and rBS $n_{\mathrm{r}}$.
		Since the index $k_{\mathrm{LoS}}$ can be obtained at step \ref{step_getkLoS} of Algorithm \ref{alg_para_esti}, 
		we reallocate the index domain $k\in \left\{ 1,\dots ,K \right\} $ to identify the $K$ targets in the following formulations.
		In addition, we will preserve the subscripts $n_{\mathrm{t}}$ and $n_{\mathrm{r}}$ in all sensing parameters to highlight the association with their respective tBS-rBS pairs, departing from the simplified notations in the previous discussions.
	\end{remark}
	
	\section{Position and Velocity Estimation}\label{Sec_Pos_Vel_Est}
	Based on the bistatic ranges $\{\hat{d}_{n_{\mathrm{t}},k,n_{\mathrm{r}}}\}$ and AoA parameters $\{\hat{\theta}_{n_{\mathrm{t}},k,n_{\mathrm{r}}},\hat{\phi}_{n_{\mathrm{t}},k,n_{\mathrm{r}}}\}$ estimated from a distributed tBS-rBS pair, the system can theoretically estimate the 3D positions of all targets. 
	However, since only bistatic Doppler velocities $\{\hat{v}_{n_{\mathrm{t}},k,n_{\mathrm{r}}}\}$ are measurable at each rBS, resolving the true 3D velocities necessitates fusing at least three independent radial velocity estimates. 
	To tackle this problem and improve the accuracy of position and velocity estimation, this section proposes a MST-based cooperative data fusion algorithm. 
	This scheme jointly resolves the data association ambiguity among estimated parameters from different tBS-rBS pairs while jointly optimizing the 3D positions and velocities of all targets.

	\subsection{Basic Target Localization}
	Let $d_{n_{\mathrm{t}},k}^{\mathrm{t}}=\left\| \mathbf{p}_{k}^{\mathrm{u}}-\mathbf{p}_{n_{\mathrm{t}}}^{\mathrm{t}} \right\| _2$ denote the distance between tBS $n_{\mathrm{t}}$ and target $k$, where $\mathbf{p}_{k}^{\mathrm{u}}=\left[ x_{k}^{\mathrm{u}},y_{k}^{\mathrm{u}},z_{k}^{\mathrm{u}} \right] ^{\mathrm{T}}$ and $\mathbf{p}_{n_{\mathrm{t}}}^{\mathrm{t}}=\left[ x_{n_{\mathrm{t}}}^{\mathrm{t}},y_{n_{\mathrm{t}}}^{\mathrm{t}},z_{n_{\mathrm{t}}}^{\mathrm{t}} \right] ^{\mathrm{T}}$ are the 3D position vectors of tBS $n_{\mathrm{t}}$ and target $k$, respectively.
	Similarly, the distance between  target $k$ and rBS $n_{\mathrm{r}}$ is defined as $d_{k,n_{\mathrm{r}}}^{\mathrm{r}}=\left\| \mathbf{p}_{k}^{\mathrm{u}}-\mathbf{p}_{n_{\mathrm{r}}}^{\mathrm{r}} \right\| _2$, where $\mathbf{p}_{n_{\mathrm{r}}}^{\mathrm{r}}=\left[ x_{n_{\mathrm{r}}}^{\mathrm{r}},y_{n_{\mathrm{r}}}^{\mathrm{r}},z_{n_{\mathrm{r}}}^{\mathrm{r}} \right] ^{\mathrm{T}}$ represents the position of rBS $n_{\mathrm{r}}$.
	Given $\hat{d}_{n_{\mathrm{t}},k,n_{\mathrm{r}}}$, $\hat{\theta}_{n_{\mathrm{t}},k,n_{\mathrm{r}}}$, and $\hat{\phi}_{n_{\mathrm{t}},k,n_{\mathrm{r}}}$, the 3D position $\mathbf{p}_{k}^{\mathrm{u}}$ of target $k$ can be determined by solving the following system of equations:
	\begin{equation}\label{Prob_basic_loca}
		\begin{cases}
			d_{n_{\mathrm{t}},k}^{\mathrm{t}}+d_{k,n_{\mathrm{r}}}^{\mathrm{r}}=\hat{d}_{n_{\mathrm{t}},k,n_{\mathrm{r}}}\\
			\mathbf{p}_{k}^{\mathrm{u}}=\mathbf{p}_{n_{\mathrm{r}}}^{\mathrm{t}}+d_{k,n_{\mathrm{r}}}^{\mathrm{r}}\hat{\mathbf{r}}^{\mathrm{r}}_{n_{\mathrm{t}},k,n_{\mathrm{r}}}\\
		\end{cases},
	\end{equation}
	where $\hat{\mathbf{r}}^{\mathrm{r}}_{n_{\mathrm{r}},k,n_{\mathrm{r}}}$ represents the unit direction vector from target $k$ to rBS $n_{\mathrm{r}}$ derived from AoA estimations, expressed in the global coordinate system as:
	\begin{align}
		&\hat{\mathbf{r}}^{\mathrm{r}}_{n_{\mathrm{t}},k,n_{\mathrm{r}}}=\mathbf{T}\left( \chi _{n_{\mathrm{r}}}^{\mathrm{r}} \right) \left[ \begin{array}{c}
			\sin \hat{\theta}_{n_{\mathrm{t}},k,n_{\mathrm{r}}}^{\mathrm{r}}\cos \hat{\phi}_{n_{\mathrm{t}},k,n_{\mathrm{r}}}^{\mathrm{r}}\\
			\sin \hat{\theta}_{n_{\mathrm{t}},k,n_{\mathrm{r}}}^{\mathrm{r}}\sin \hat{\phi}_{n_{\mathrm{t}},k,n_{\mathrm{r}}}^{\mathrm{r}}\\
			\cos \hat{\phi}_{n_{\mathrm{t}},k,n_{\mathrm{r}}}^{\mathrm{r}}\\
		\end{array} \right] ,
		\\
		&\mathbf{T}\left( \chi _{n_{\mathrm{r}}}^{\mathrm{r}} \right) =\left[ \begin{matrix}
			\cos \chi _{n_{\mathrm{r}}}^{\mathrm{r}}&		\sin \chi _{n_{\mathrm{r}}}^{\mathrm{r}}&		0\\
			-\sin \chi _{n_{\mathrm{r}}}^{\mathrm{r}}&		\cos \chi _{n_{\mathrm{r}}}^{\mathrm{r}}&		0\\
			0&		0&		1\\
		\end{matrix} \right],
	\end{align}
	where $\chi _{n_{\mathrm{r}}}^{\mathrm{r}}$ is the horizontal orientation of rBS $n_{\mathrm{r}}$'s array, defined as the angle between the array normal and the global $x$-axis.
	Solving \eqref{Prob_basic_loca} for $\forall k$ yields basic position estimates $\{ \hat{\mathbf{p}}_{n_{\mathrm{t}},k,n_{\mathrm{r}}}^{\mathrm{u}} \} _{k=1}^{K}$ related to tBS $n_{\mathrm{t}}$ and rBS $n_{\mathrm{r}}$.
	However, independent processing across the tBS-rBS pairs introduces data association ambiguity: 
	due to inconsistent indexing across pairs,
	the estimated $\hat{\mathbf{p}}_{n_{\mathrm{t}},k,n_{\mathrm{r}}}^{\mathrm{u}}$ and $\hat{\mathbf{p}}_{n^\prime _{\mathrm{t}},k,n^\prime _{\mathrm{r}}}^{\mathrm{u}}$ may correspond to different targets when $n_{\mathrm{t} }\ne n^\prime _{\mathrm{t}}$ or $n _{\mathrm{r}} \ne n^\prime _{\mathrm{r}}$.
	In addition, estimation errors in $\hat{d}_{n_{\mathrm{t}},k,n_{\mathrm{r}}}$, $\hat{\theta}_{n_{\mathrm{t}},k,n_{\mathrm{r}}}$, and $\hat{\phi}_{n_{\mathrm{t}},k,n_{\mathrm{r}}}$ lead to biases in position estimates.
	Although greedy algorithms can be applied for the target-to-estimate association \cite{9583869}, the erroneous position estimates cannot be effectively eliminated. 
	This limitation often leads to incorrect data associations or significant global positioning biases, particularly in scenarios with a large number of targets or low SNRs.
	In the following, an effective error elimination and data association algorithm is proposed based on the MST method.

	\subsection{Error Elimination and Data Association}\label{Sec_DataAssociation}
\begin{figure}[!t]
		\centering
		\includegraphics[width=0.9\linewidth]{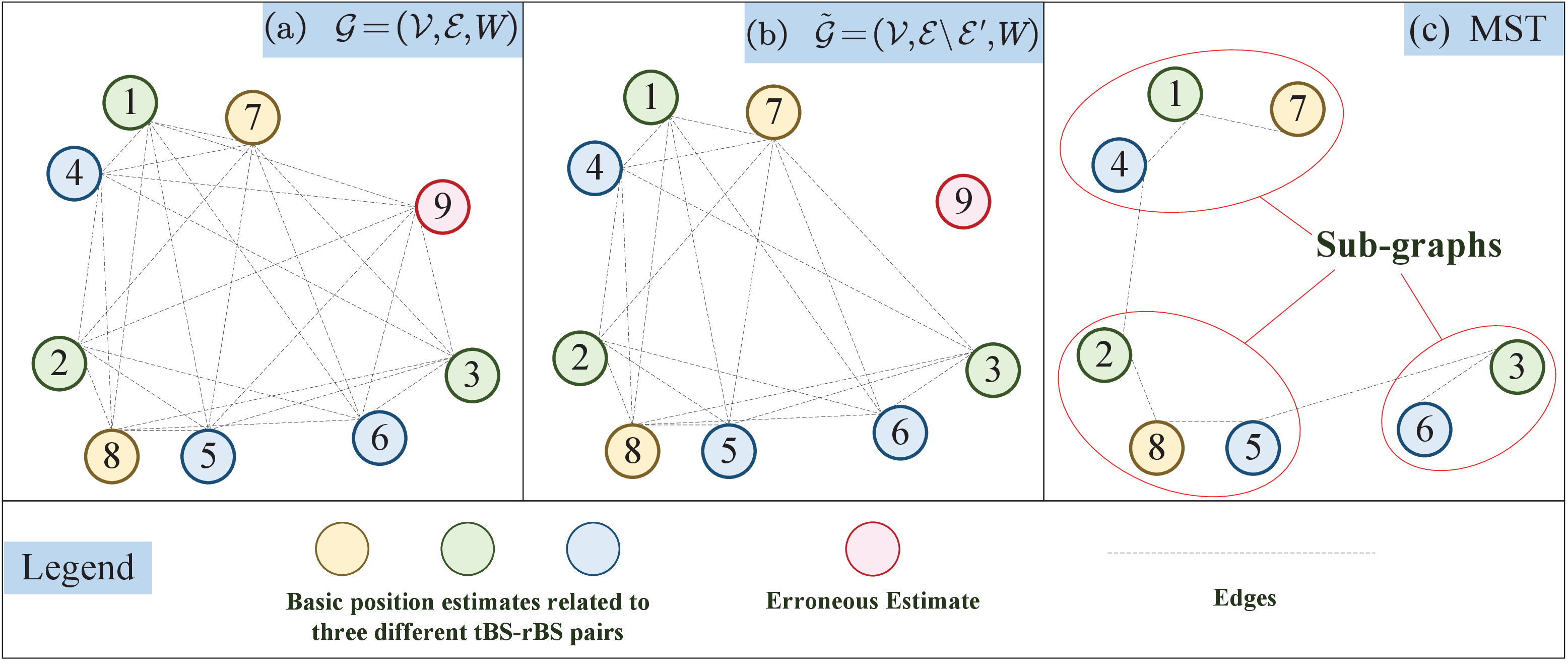}
		\caption{An illustration of the MST-based error elimination and data association process, where the numbers of targets and tBS-rBS pairs are $K=3$ and $N_{\mathrm{t}} N_{\mathrm{r}}=3$, receptively.}
		\label{fig:MST}
		\vspace{-0.3cm}
	\end{figure}

	To establish the MST-based data association framework, we first define an undirected weighted graph $\mathcal{G} =\left( \mathcal{V} ,\mathcal{E} ,W\right) $ consisting of the vertex set $\mathcal{V} $, the edge set $\mathcal{E}$, and the weight function $W$ as illustrated in Fig. \ref{fig:MST}(a).
	Specifically, $\mathcal{V} $ is represented as a union of $N_{\mathrm{t}}\times N_{\mathrm{r}}$ subsets as follows
	\begin{equation}\label{equation:Graph_V}
		\mathcal{V} =\bigcup_{n_{\mathrm{t}}=1}^{N_{\mathrm{t}}}{\bigcup_{n_{\mathrm{r}}=1}^{N_{\mathrm{r}}}{\mathcal{V} _{n_{\mathrm{t}},n_{\mathrm{r}}}}},
	\end{equation}
	where each subset $\mathcal{V} _{n_{\mathrm{t}},n_{\mathrm{r}}}$ corresponds to the estimates from tBS $n_{\mathrm{t}}$ and rBS $n_{\mathrm{r}}$, indexed as
	\begin{equation}
		\mathcal{V} _{n_{\mathrm{t}},n_{\mathrm{r}}}= \{ ( ( n_{\mathrm{r}}-1 ) N_{\mathrm{t}}+n_{\mathrm{t}}-1 ) K+1,\dots ,( ( n_{\mathrm{r}}-1 ) N_{\mathrm{t}}+n_{\mathrm{t}} ) K \}.
	\end{equation}
	The edge $\mathcal{E}$ connects vertices across different tBS-rBS pairs, defined as
	\begin{equation}
		\mathcal{E} =\left\{ \left( e_1,e_2 \right) |e_1\in V_{n_{\mathrm{t}},n_{\mathrm{r}}},\, e_2\in V\backslash V_{n_{\mathrm{t}},n_{\mathrm{r}}} \right\} .
	\end{equation}
	The edge weight between any two vertices $e_1$ and $e_2$ is defined as the Euclidean distance between their corresponding position estimates as follows
	\begin{align}\label{equation:Graph_W}
		&W\left( e_1,e_2 \right) =\left\| \hat{\mathbf{p}}_{\xi ^{\mathrm{t}}\left( e_1 \right) ,\xi ^{\mathrm{u}}\left( e_1 \right) ,\xi ^{\mathrm{r}}\left( e_1 \right)}-\hat{\mathbf{p}}_{\xi ^{\mathrm{t}}\left( e_2 \right) ,\xi ^{\mathrm{u}}\left( e_2 \right) ,\xi ^{\mathrm{r}}\left( e_2 \right)} \right\| _2, \notag
		\\
		&\forall \left( e_1,e_2 \right) \in \mathcal{E} ,
	\end{align}
	where $\xi ^{\mathrm{t}}\left( e \right) =\mathrm{mod}\left( \mathrm{mod}\left( e-1,K \right) ,N_{\mathrm{t}} \right) +1$, $\xi ^{\mathrm{u}}\left( e \right) =\lfloor \frac{e_1-1}{K} \rfloor +1$ and $\xi ^{\mathrm{r}}\left( e \right) =\lfloor \frac{\mathrm{mod}\left( e_1-1,K \right)}{N_{\mathrm{t}}} \rfloor +1$ decodes the tBS-target-rBS indices from vertex $e$, respectively,
	with $\mathrm{mod}$ and $\lfloor \cdot \rfloor $ denoting the modulus and floor operations, respectively.
	

	It can be noted that valid position estimates of the same target should cluster spatially, while erroneous estimates deviate significantly.
	That is, a vertex is likely corresponding to erroneous estimate if all its adjacent edges have large weights.
	To eliminate such outliers, we introduce a threshold $\varpi$ and define the following pruned edge set
	\begin{equation}
		\mathcal{E} ^{\prime}=\left\{ \left( e_1,e_2 \right)\in \mathcal{E} |W_{\min}\left( e_1 \right) >\varpi  \right\} ,
	\end{equation}
	where $W_{\min}\left( e_1 \right)$ denotes the minimum weight among all edges adjacent to $e_1$.
	By removing these edges, the graph $\mathcal{G} $ is updated as
	\begin{equation}\label{equation:updateG}
		\tilde{\mathcal{G}}=\left( \mathcal{V} , \mathcal{E} \backslash \mathcal{E} ^{\prime} , W\right) ,
	\end{equation}
	which isolates vertices corresponding to erroneous estimates as illustrated in Fig. \ref{fig:MST}b.
	
	Finally, the MST of the pruned graph $\tilde{\mathcal{G}}$ can be computed using the well-established algorithms such as Prim's \cite{PrimShortest} or Kruskal's \cite{KruskalOn}.
	Since the valid position estimates are clustered, the edges connecting  estimates of the same target usually exhibit smaller weights and are preferentially retained in the MST.
	Therefore, the final data association is achieved by removing the $K-1$ longest edges from the MST.
	This operation partitions the MST into $K$ sub-graphs as shown in Fig. \ref{fig:MST}(c), each corresponding to a unique target.

	\subsection{Data Fusion for Target Localization}
	Based on the MST-based association results in the previous subsection, we develop a data fusion scheme for cooperative target localization. 
	While the subsequent derivations retain summation over all sensing parameters for notational continuity, we emphasize that only the validated parameters (after error elimination) contribute to the final estimation.
	
	The fused position estimates $\left\{ \accentset{*}{\mathbf{p}}_{k}^{\mathrm{u}} \right\} _{k=1}^{K}$ can be obtained through either hard or soft fusion \cite{Han2024}.
	In the hard fusion approach, basic estimates $\{\hat{\mathbf{p}}_{n_{\mathrm{t}},k,n_{\mathrm{r}}}^{\mathrm{u}}\}$ are directly combined, for instance via arithmetic averaging:
	\begin{equation} \label{equation:average_position}
		\accentset{*}{\mathbf{p}}_{k}^{\mathrm{u},\mathrm{ave}}=\frac{1}{N_{\mathrm{t}}N_{\mathrm{r}}}\sum_{n_{\mathrm{t}}=1}^{N_{\mathrm{t}}}{\sum_{n_{\mathrm{r}}=1}^{N_{\mathrm{r}}}{\hat{\mathbf{p}}_{n_{\mathrm{t}},k,n_{\mathrm{r}}}^{\mathrm{u}}}}.
	\end{equation}
	To achieve higher localization accuracy, we adopt a soft fusion paradigm by formulating a maximum-likelihood estimator that jointly processes all original  sensing parameters $\{\hat{d}_{n_{\mathrm{t}},k,n_{\mathrm{r}}},\hat{\theta}_{n_{\mathrm{t}},k,n_{\mathrm{r}}},\hat{\phi}_{n_{\mathrm{t}},k,n_{\mathrm{r}}}\}$.
	Specifically, we define the following loss function
	\begin{align}
		f\left( \mathbf{p}_{k}^{\mathrm{u}} \right) =&\frac{\sum_{n_{\mathrm{t}}=1}^{N_{\mathrm{t}}}{\sum_{n_{\mathrm{r}}=1}^{N_{\mathrm{r}}}{\alpha _{n_{\mathrm{t}},k,n_{\mathrm{r}}}\left| \hat{d}_{n_{\mathrm{t}},k,n_{\mathrm{r}}}-d_{n_{\mathrm{t}},k}^{\mathrm{t}}-d_{k,n_{\mathrm{r}}}^{\mathrm{r}} \right|}}}{\sum_{n_{\mathrm{t}}=1}^{N_{\mathrm{t}}}{\sum_{n_{\mathrm{r}}=1}^{N_{\mathrm{r}}}{\alpha _{n_{\mathrm{t}},k,n_{\mathrm{r}}}}}}\notag
		\\
		&+\frac{\sum_{n_{\mathrm{t}}=1}^{N_{\mathrm{t}}}{\sum_{n_{\mathrm{r}}=1}^{N_{\mathrm{r}}}{\beta _{n_{\mathrm{t}},k,n_{\mathrm{r}}}\left\| \hat{\mathbf{r}}^{\mathrm{r}}_{n_{\mathrm{r}},k,n_{\mathrm{r}}}-\mathbf{r}^{\mathrm{r}}_{k,n_{\mathrm{r}}} \right\|}}}{\sum_{n_{\mathrm{t}}=1}^{N_{\mathrm{t}}}{\sum_{n_{\mathrm{r}}=1}^{N_{\mathrm{r}}}{\beta _{n_{\mathrm{t}},k,n_{\mathrm{r}}}}}},
	\end{align}
	where $\mathbf{r}_{k,n_{\mathtt{r}}}^{\mathtt{r}}=\left( \mathbf{p}_{n_{\mathtt{r}}}^{\mathtt{r}}-\mathbf{p}_{k}^{\mathtt{u}} \right) /\left\| \mathbf{p}_{n_{\mathtt{r}}}^{\mathtt{r}}-\mathbf{p}_{k}^{\mathtt{u}} \right\| _2$ denotes the unit direction vector from target $k$ to rBS $n_{\mathrm{r}}$ determined by the target position $\mathbf{p}_{k}^{\mathrm{u}}$.
	The weights $\alpha _{n_{\mathrm{t}},k,n_{\mathrm{r}}}=( d_{n_{\mathrm{t}},k}^{\mathrm{t}}d_{k,n_{\mathrm{r}}}^{\mathrm{r}} ) ^{-2}$ and $\beta _{n_{\mathrm{t}},k,n_{\mathrm{r}}}=\alpha _{n_{\mathrm{t}},k,n_{\mathrm{r}}}d_{k,n_{\mathrm{t}}}^{\mathrm{r}}$ are designed based on the fact that the signal path loss is inversely proportional to $\left( d_{n_{\mathrm{t}}}^{\mathrm{t}}\left( \mathbf{p}_{k}^{\mathrm{u}} \right) \cdot d_{n_{\mathrm{r}}}^{\mathrm{r}}\left( \mathbf{p}_{k}^{\mathrm{u}} \right) \right) ^{2}$, thereby assigning higher confidence to sensing parameters derived under stronger SNRs \cite{Han2024}.
	Then, the optimal position estimate $\accentset{*}{\mathbf{p}}_{k}^{\mathrm{u}}$ is obtained by solving the following problem
	\begin{equation}\label{Prob_PosEsti}
		\min_{\mathbf{p}_{k}^{\mathrm{u}}} f\left( \mathbf{p}_{k}^{\mathrm{u}} \right) .
	\end{equation}
	Problem \eqref{Prob_PosEsti} is an unconstrained optimization problem, which can be efficiently solved by the quasi-Newton algorithm \cite{GillQuasi}.
	In this work, we initialize the quasi-Newton solver with the averaging-based estimate $\accentset{*}{\mathbf{p}}_{k}^{\mathrm{u},\mathrm{ave}}$, leveraging its proximity to the true position to mitigate the sensitivity of non-convex optimization to initial conditions.
	
	\vspace{-0.1cm}
	\subsection{Velocity Estimation}
	Leveraging the cooperative position estimates obtained in the previous subsection, we estimate the true 3D velocity $\mathbf{v}_k=\left[ v_x,v_y,v_z \right] ^{\mathrm{T}}$ of target $k$ using its bistatic Doppler velocity estimates $\{\hat{v}_{n_{\mathrm{t}},k,n_{\mathrm{r}}}|\forall n_{\mathrm{t}},n_{\mathrm{r}}\}$.
	Here, the sensing parameters $d_{n_{\mathrm{t}},k}^{\mathrm{t}}$, $d_{k,n_{\mathrm{r}}}^{\mathrm{r}}$, $\mathbf{r}_{k,n_{\mathrm{r}}}^{\mathrm{r}}$, and $ \alpha _{n_{\mathrm{t}},k,n_{\mathrm{r}}}$ are recalculated using $\accentset{*}{\mathbf{p}}_{k}^{\mathrm{u}}$ instead of the unknown true position $\mathbf{p}_{k}^{\mathrm{u}}$
	, with $\mathbf{r}_{n_{\mathrm{t}},k}^{\mathrm{t}}$ denoting the unit direction vector from target $k$ to tBS $n_{\mathrm{t}}$.
	
	The velocity estimation can be formulated as the following weighted least squares problem
	\begin{equation} \label{Prob_Esti_v}
		\!\!\min_{\mathbf{v}_k}\! \sum_{n_{\mathrm{t}}=1}^{N_{\mathrm{t}}}{\sum_{n_{\mathrm{r}}=1}^{N_{\mathrm{r}}}\!{w_{n_{\mathrm{t}},k,n_{\mathrm{r}}}\!\left( \hat{v}_{n_{\mathrm{t}},k,n_{\mathrm{r}}}-\left( \mathbf{r}_{n_{\mathrm{t}},k}^{\mathrm{t}}+\mathbf{r}_{k,n_{\mathrm{r}}}^{\mathrm{r}} \right) ^{\mathrm{\!T}}\mathbf{v}_k \right)\! ^2,}}\!
	\end{equation}
	where the weights $w_{n_{\mathrm{t}},k,n_{\mathrm{r}}}$ inherit the path-loss-aware scaling from the position estimation:
	\begin{equation}
		w_{n_{\mathrm{t}},k,n_{\mathrm{r}}}=\frac{\alpha _{n_{\mathrm{t}},k,n_{\mathrm{r}}}}{\sum_{n_{\mathrm{t}}=1}^{N_{\mathrm{t}}}{\sum_{n_{\mathrm{r}}=1}^{N_{\mathrm{r}}}{\alpha _{n_{\mathrm{t}},k,n_{\mathrm{r}}}}}}.
	\end{equation}
	To solve this problem efficiently, we vectorize it as 
	\begin{equation} \label{Prob_Esti_v_vectorize}
		\min_{\mathbf{v}_k} \left( \hat{\mathbf{v}}_k-\hat{\mathbf{R}}_k\mathbf{v}_k \right) ^{\mathrm{T}}\mathbf{W}_k\left( \hat{\mathbf{v}}_k-\hat{\mathbf{R}}_k\mathbf{v}_k \right) ,
	\end{equation}
	where $\hat{\mathbf{v}}_k\in \mathbb{R} ^{N_{\mathrm{t}}N_{\mathrm{r}}\times 1}$, $\hat{\mathbf{R}}_k\in \mathbb{R} ^{N_{\mathrm{t}}N_{\mathrm{r}}\times 3}$, and $\mathbf{W}_k\in \mathbb{R} ^{N_{\mathrm{t}}N_{\mathrm{r}}\times N_{\mathrm{t}}N_{\mathrm{r}}}$ are constructed as
	\begin{align}
		&\hat{\mathbf{v}}_k=\left[ \hat{v}_{1,k,1},\dots ,\hat{v}_{N_{\mathrm{t}}k,N_{\mathrm{r}}} \right] ^{\mathrm{T}},
		\\
		&\hat{\mathbf{R}}_k=\left[ \left( \mathbf{r}_{1,k}^{\mathrm{t}}+\mathbf{r}_{k,1}^{\mathrm{r}} \right) ,\dots ,\left( \mathbf{r}_{N_{\mathrm{t}},k}^{\mathrm{t}}+\mathbf{r}_{k,N_{\mathrm{r}}}^{\mathrm{r}} \right) \right] ^{\mathrm{T}},
		\\
		&\mathbf{W}_k=\mathrm{diag}\left( \left[ w_{1,k,1},\dots ,w_{N_{\mathrm{t}},k,N_{\mathrm{r}}} \right] \right) .
	\end{align}
	Recall that $N_{\mathrm{t}}N_{\mathrm{r}}\geqslant 3$ independent radial velocity estimates are required for 3D velocity estimation, which means $\mathrm{rank}( \hat{\mathbf{R}}_k ) =3$.
	Hence, the closed-form optimal solution of Problem \eqref{Prob_Esti_v_vectorize} is given by
	\begin{equation}\label{equation:cal_vk}
		\accentset{*}{\mathbf{v}}_k=\left( \hat{\mathbf{R}}_{k}^{\mathrm{T}}\mathbf{W}_k\hat{\mathbf{R}}_k \right) ^{-1}\hat{\mathbf{R}}_{k}^{\mathrm{T}}\mathbf{W}_k\hat{\mathbf{v}}_k.
	\end{equation}
	
	\vspace{-0.2cm}
	\subsection{Algorithm Development}
	
	\begin{algorithm}[t]
		\caption{Cooperative Position and Velocity Estimation}
		\label{alg_pos_vel_Esti}
		\begin{algorithmic}[1] 
			\Require 
			Sensing parameters $\{\hat{d}_{n_{\mathrm{t}},k,n_{\mathrm{r}}},\hat{\theta}_{n_{\mathrm{t}},k,n_{\mathrm{r}}},\hat{\phi}_{n_{\mathrm{t}},k,n_{\mathrm{r}}}\hat{v}_{n_{\mathrm{t}},k,n_{\mathrm{r}}}\}$.
			\Ensure 
			3D position and velocity estimates of targets $\{\accentset{*}{\mathbf{p}}_{k}^{\mathrm{u}},\accentset{*}{\mathbf{v}}_k\}$.
			\For{$\forall n_{\mathrm{t}},k,n_{\mathrm{r}}$} 
			\Comment{\textbf{Basic position estimation}}
			\State Obtain $\hat{\mathbf{p}}_{n_{\mathrm{t}},k,n_{\mathrm{r}}}^{\mathrm{u}}$ by solving system of equations \eqref{Prob_basic_loca};
			\EndFor
			\State Construct undirected weighted graph $\mathcal{G} =\left( \mathcal{V} ,\mathcal{E} ,W\right) $ via \eqref{equation:Graph_V}-\eqref{equation:Graph_W};
			\Comment{\textbf{Data association}}
			\State Obtain graph $\tilde{\mathcal{G}}$ via \eqref{equation:updateG};
			\State Generate the MST of $\tilde{\mathcal{G}}$;\label{step_MST}
			\State Obtain sub-graphs by removing the $K-1$ longest edges from the MST;
			\State Obtain data association results from vertices in each subgraph;
			\For{$\forall k$} 
			\Comment{\textbf{Cooperative position estimation}}
			\State Obtain $\accentset{*}{\mathbf{p}}_{k}^{\mathrm{u}}$ by solving Problem \eqref{Prob_PosEsti};\label{step_Prob_PosEsti}
			\EndFor
			\For{$\forall k$} 
			\Comment{\textbf{Velocity estimation}}
			\State Calculate $\accentset{*}{\mathbf{v}}_k$ via \eqref{equation:cal_vk}; \label{cal_vk}
			\EndFor
		\end{algorithmic}
	\end{algorithm}
	
	Based on the above discussions, we summarize the proposed cooperative position and velocity estimation algorithm in Algorithm \ref{alg_pos_vel_Esti}.
	The algorithm comprises four cascaded phases: 1) basic position estimation, 2) MST-based data association, 3) cooperative position optimization, and 4) velocity estimation.
	
	\textit{Complexity Analysis:}
	The computational complexity of Algorithm \ref{alg_pos_vel_Esti} is dominated by the MST generation in Step \ref{step_MST} and solving Problem \eqref{Prob_PosEsti} in Step \ref{step_Prob_PosEsti}.
	The MST generation via Kruskal's algorithm incurs a complexity of $O\left( \left| \mathcal{E} \backslash \mathcal{E} ^{\prime} \right|\log \left| \mathcal{E} \backslash \mathcal{E} ^{\prime} \right| \right)$, where 
	$\left| \mathcal{E} \backslash \mathcal{E} ^{\prime} \right|\leqslant \left| \mathcal{E} \right|<\left( KN_{\mathrm{t}}N_{\mathrm{r}} \right) ^2$.
	Given typical ISAC network scales, this simplifies to $O( \left( KN_{\mathrm{t}}N_{\mathrm{r}} \right) ^2\log \left( KN_{\mathrm{t}}N_{\mathrm{r}} \right) )$.
	To solve Problem \eqref{Prob_PosEsti}, the computational complexity of each quasi-Newton iteration is of $O\left( N_{\mathrm{t}}N_{\mathrm{r}} \right) $.
	With the averaging-based initialization $\accentset{*}{\mathbf{p}}_{k}^{\mathrm{u}}$ in \eqref{equation:average_position}, the required iteration number $G^{\prime}$ remains small (typically $G^{\prime}\leqslant 5$).
	Therefore, the total complexity of generating the fused position estimates $\left\{ \accentset{*}{\mathbf{p}}_{k}^{\mathrm{u}} \right\} _{k=1}^{K}$ is of $O\left( KG^{\prime}N_{\mathrm{t}}N_{\mathrm{r}} \right) $.
	To sum up, the total complexity of Algorithm \ref{alg_pos_vel_Esti} is given by $O( ( KN_{\mathrm{t}}N_{\mathrm{r}} ) ^2\log \left( KN_{\mathrm{t}}N_{\mathrm{r}} \right) +KG^{\prime}N_{\mathrm{t}}N_{\mathrm{r}} ) $.
	
	Notably, the computational overhead of data fusion for localization and velocity estimation is negligible compared to the parameter extraction phase, which ensures the scalability of cooperative ISAC systems even with increasing network density.

	\section{Simulation Results}\label{Sec_Simu_Result}
	In this section, numerical examples are presented to investigate the performance of the proposed MIMO-OFDM cellular network-based cooperative bistatic sensing framework.
	
	\subsection{Simulation Setups}
\begin{figure}
		\centering
		\includegraphics[width=0.6\linewidth]{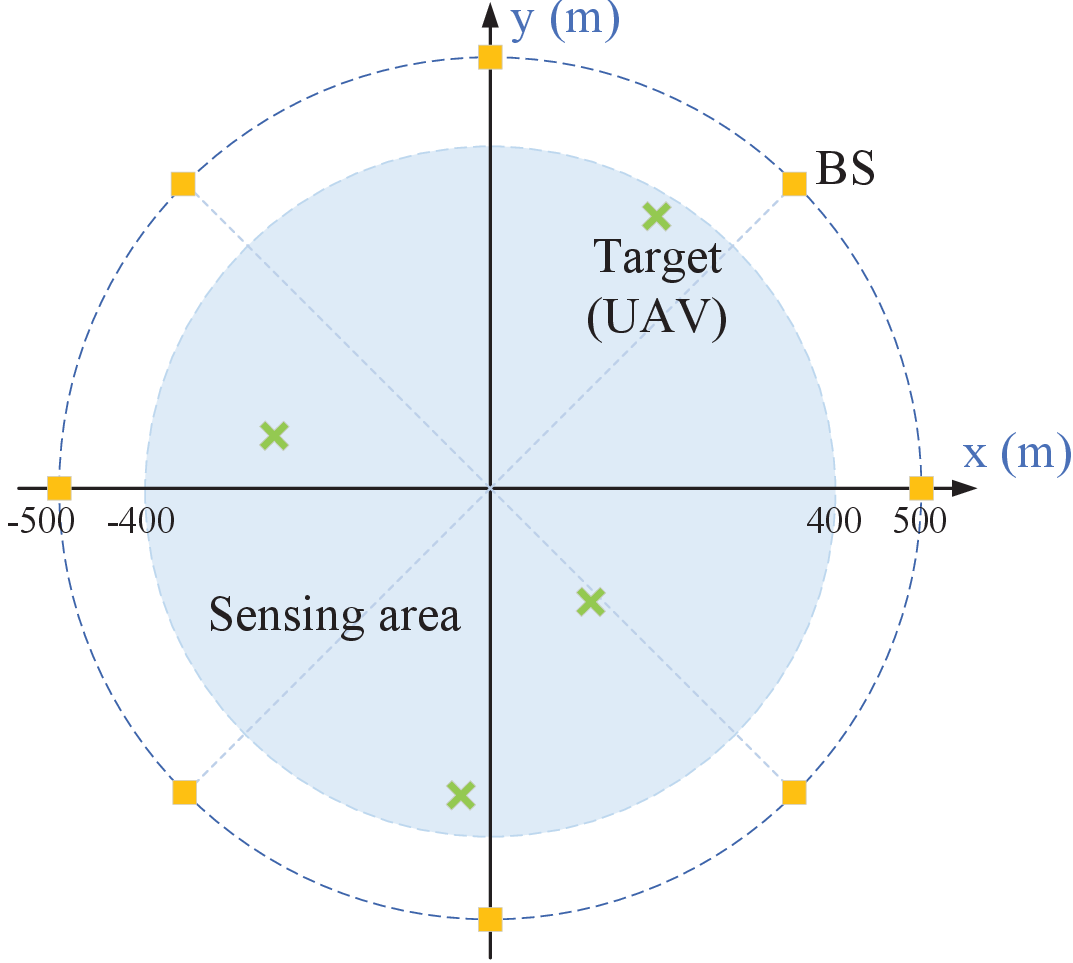}
		\caption{Illustration of the system model in simulation.}
		\vspace{-0.2cm}
		\label{fig:SimuModel}
	\end{figure}

	The simulation considers a cooperative ISAC system comprising 8 BSs uniformly distributed on a circle with a radius of 500 meters (m), as illustrated in Fig. \ref{fig:SimuModel}. 
	Each BS is positioned at a height of 30 m and equipped with a uniform planar array (UPA) consisting of 16 horizontal and 24 vertical antenna elements, driven by 64 radio frequency (RF) chains. 
	The horizontal orientation $\chi _{n_{\mathrm{r}}}^{\mathrm{r}}$ of each UPA is configured to point towards the center of the circle. 
	The precoding matrices $\mathbf{F}$ and the combining matrices $\mathbf{Q}$ for all BSs are constructed following the exemplary design in Section \ref{Sec_Beanforming_LAE}.

	UAVs are modelled as sensing targets, generated uniformly at random within a circular area of radius 400 m and at heights ranging between 50 m and 300 m. 
	The UAVs exhibit a maximum flight speed of 60 km/h and a minimum speed of 5 km/h, with a safety separation distance of 10 m enforced to prevent collisions. 
	The RCS of each UAV is set to $\sigma _{\mathrm{RCS}} = 0.01$ m\textsuperscript{2}.
	
	Sensing signal transmission utilizes a typical sub-6 GHz configuration with a carrier frequency of $f_{\mathrm{c}}=4.9$ GHz and a subcarrier spacing of $\Delta f=30$ kHz.
	Unless otherwise stated, each tBS occupies a BWP of 20 MHz (including guard intervals), transmitting sensing signals over 51 resource blocks (RBs), where each RB comprises 12 subcarriers \cite{3GPP13810101}. 
	Each sensing task employs $N = 7$ consecutive OFDM symbols.
	The path losses for the target-scattered paths and the LoS paths are respectively modelled as follows \cite{ITURRECP525}
	\begin{align}
		&\mathrm{PL}_{\mathrm{Ts}}=103.4+20\lg f_{\mathrm{c}}+20\lg d^{\mathrm{t}}+20\lg d^{\mathrm{r}}-10\lg \sigma _{\mathrm{RCS}},
		\\
		&\mathrm{PL}_{\mathrm{LoS}}=32.4+20\lg f_{\mathrm{c}}+20\lg d^{\mathrm{LoS}},
	\end{align}
	where $d^{\mathrm{t}}$ and $d^{\mathrm{r}}$ are the distances in km from the target to the tBS and rBS, respectively, and $d^{\mathrm{LoS}}$ denotes the distance in km between the tBS and rBS.
	The STO and CFO between each tBS-rBS pair are set to $10^{-8}$ s and $0.01 \Delta f$, respectively.

	The root mean square error (RMSE) is employed to quantify the sensing accuracy in the simulations. 
	For any sensing quantity $\mathbf{x}$, which may represent bistatic ranges, Doppler velocities, AoA estimates, or 3D position/velocity estimates, the RMSE is defined as:
	\begin{equation}
		\mathrm{RMSE}\left( \mathbf{x} \right) = \sqrt{\frac{1}{K}\sum_{k=1}^K{\left\| \hat{\mathbf{x}}_k - \mathbf{x}_k \right\| _{2}^{2}}},
	\end{equation}
	where $\hat{\mathbf{x}}_k$ and $\mathbf{x}_k$ denote the estimated and true value for target $k$, respectively.
	All simulation results are derived from 500 independent Monte Carlo trials.
	To mitigate the influence of outliers, the reported results represent the average over the top 95\% of samples with the highest estimation accuracy, excluding the bottom 5\%.
	Note that all simulation results are presented on a logarithmic scale for the y-axis.
	
	\subsection{Sensing Parameter Extraction Performance}
	Recall that all rBSs in the proposed sensing framework execute identical parameter extraction algorithms. 
	To evaluate the parameter extraction performance of the proposed scheme, we first consider a bistatic sensing scenario with a single tBS-rBS pair deployed at (-500 m, 0) and (500 m, 0).
	The core challenge in estimating bistatic ranges $\left\{ d_k \right\}$, azimuth/elevation angles $\left\{ \theta _{k}^{\mathrm{r}},\phi _{k}^{\mathrm{r}} \right\}$, and bistatic Doppler velocities $\left\{ v_k \right\} $ is to recover the factor matrices $\mathbf{A}$, $\mathbf{B}$, and $\mathbf{C}$ by solving Problem \eqref{Proble_CP_decomp}.
	For performance benchmarking, we introduce a baseline scheme termed \textbf{ALS}\footnote{Although other parameter extraction algorithms such as 2D-MUSIC or 2D-FFT exist, they estimate each set of parameters separately and cannot directly match parameters belonging to the same target in multi-target scenarios. ALS is adopted as it is the only conventional method capable of jointly estimating all parameters with correct pairing.}.
	This approach solves Problem \eqref{Proble_CP_decomp} using the ALS algorithm \cite{zhouzhou,C-DRCNN}, serving as a comparative reference to the proposed Algorithm \ref{alg_para_esti}.
	
	\begin{figure}
		\centering
		\includegraphics[width=0.6\linewidth]{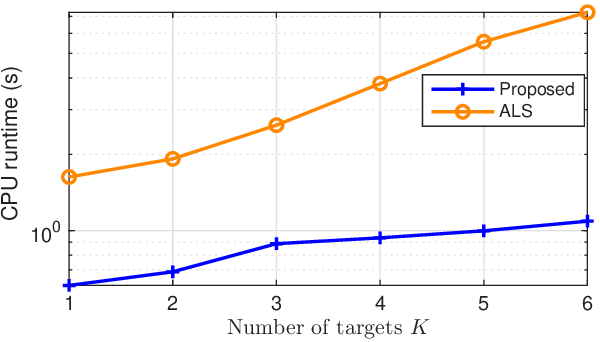}
		\caption{CPU time versus the number of targets $K$.}
		\vspace{-0.2cm}
		\label{fig:CPUtime}
	\end{figure}
	
	Fig. \ref{fig:CPUtime} compares the computational complexity of the proposed scheme and the ALS baseline, measured by CPU runtime versus the number of sensing targets $K$.
	The proposed algorithm demonstrates excellent computational efficiency, with its runtime showing almost negligible growth when the number of targets exceeds $K=3$.
	On the contrary, the ALS scheme requires higher computational resources in all scenarios, with its runtime increasing approximately exponentially as $K$ grows.
	Moreover, the computational efficiency of the proposed algorithm is not significatly affected by the growth in the number of targets.
	This indicates the scalability advantage of the proposed algorithm in real-time multi-target sensing applications, particularly in dense environments where system implementation is constrained by computational resources.

	\begin{figure}[tbp]
		\centering
		\begin{subfigure}{0.48\columnwidth}
			\centering
			\includegraphics[width=\linewidth]{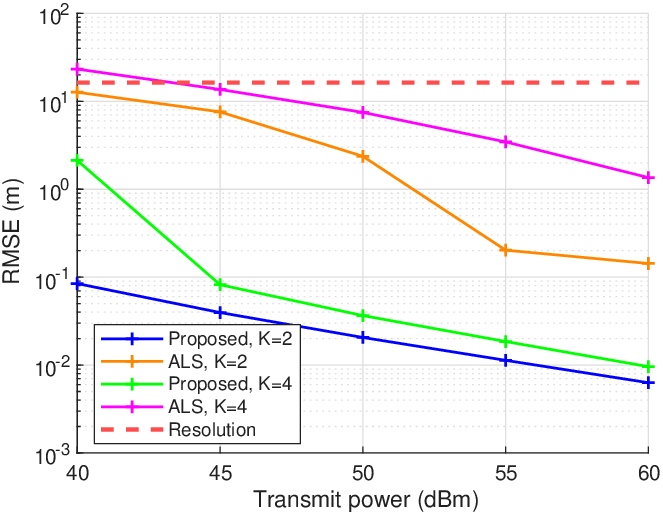} 
			\vspace{-0.4cm}
			\caption{Bistatic range}
			\label{fig:RMSE_range}
		\end{subfigure}
		\hfill 
		\begin{subfigure}{0.48\columnwidth}
			\centering
			\includegraphics[width=\linewidth]{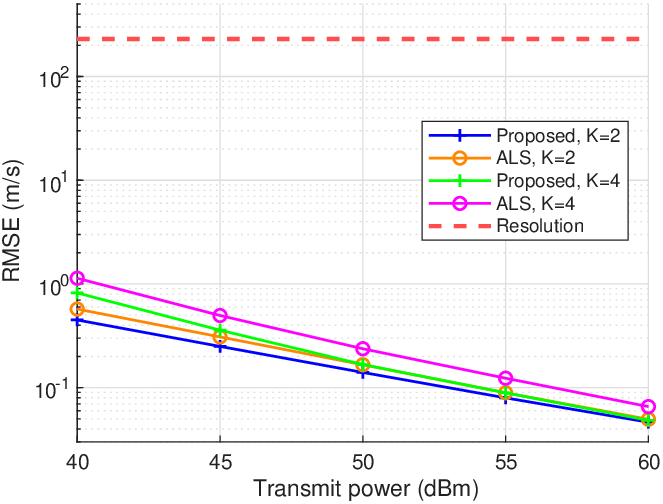} 
			\vspace{-0.4cm}
			\caption{Bistatic Doppler velocity}
			\label{fig:RMSE_velocity}
		\end{subfigure}
		\vspace{0.5\baselineskip} 
		
		\begin{subfigure}{0.48\columnwidth}
			\centering
			\includegraphics[width=\linewidth]{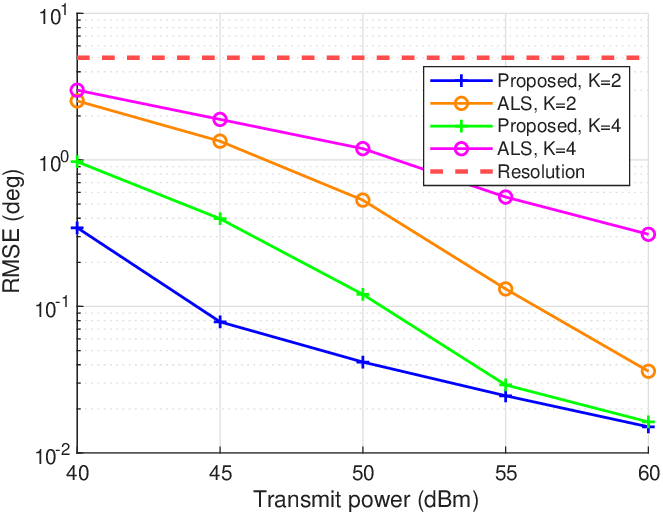}
			\vspace{-0.4cm}
			\caption{Elevation angle}
			\label{fig:RMSE_elevationAoA}
		\end{subfigure}
		\hfill 
		\begin{subfigure}{0.48\columnwidth}
			\centering
			\includegraphics[width=\linewidth]{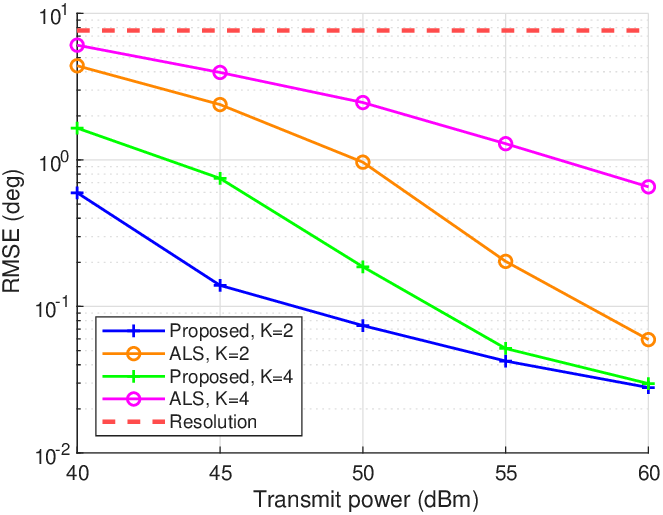}
			\vspace{-0.4cm}
			\caption{Azimuth angle}
			\label{fig:RMSE_azimuthAoA}
		\end{subfigure}
		
		\caption{Parameter estimation accuracy versus transmit power.} 
		\label{fig:Para_extract}
	\end{figure}
	
	Fig. \ref{fig:Para_extract} presents the parameter estimation RMSE performance for $K=\left\lbrace 2,4\right\rbrace $ targets under varying transmit power at the tBS.
	In each subfigure, a horizontal line is included to indicate the theoretical resolution limit for the corresponding sensing parameter, which is determined by the system configuration.		
	It can be observed that both schemes exhibit monotonic improvements in estimation accuracy with increasing tBS transmit power, resulting from the increment of SNR.
	The proposed algorithm demonstrates significant advantages in estimation accuracy across all parameters, particularly in bistatic range estimation, where it outperforms the baseline by one to two orders of magnitude, achieving millimeter-level precision under sufficient transmit power.
	Conversely, the ALS scheme shows limited estimation accuracy due to slow and unstable convergence.
	Interestingly, thanks to the joint extraction of time delays, Doppler shifts, and angles from the received signal tensor $\mathcalbf{Y}$, the proposed scheme achieves an RMSE far below the theoretical resolution limit, especially in Doppler velocity estimation.
	It verifies that the proposed algorithm achieves considerable sensing accuracy with only a small number of OFDM symbols ($N=7$), thereby minimizing the impact on the system's communication capacity.
	
\begin{figure}[tbp]
		\centering
		\includegraphics[width=0.65\linewidth]{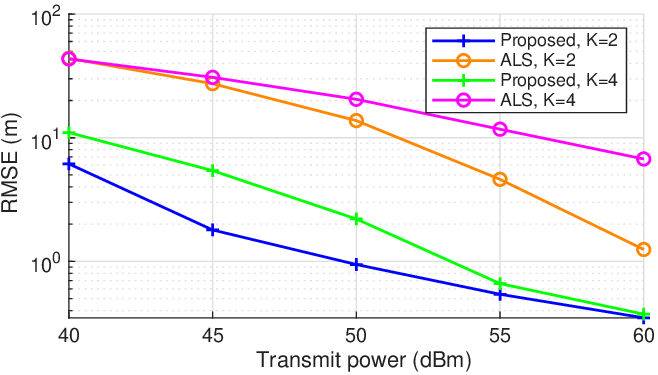}
		\caption{Basic localization accuracy versus transmit power.}
		\vspace{-0.2cm}
		\label{fig:RMSE_posiRMSE_Basic}
	\end{figure}

	Based on these parameter estimates, Fig. \ref{fig:RMSE_posiRMSE_Basic} evaluates the RMSE of the basic target position estimation derived from \eqref{Prob_basic_loca}.
	Thanks to the high-precision parameter extraction ability of Algorithm \ref{alg_para_esti}, the proposed scheme achieves a localization accuracy approximately one order of magnitude higher than the ALS scheme, with a decimetre-level positioning accuracy when the transmit power reaches 55 dBm.
	
	\vspace{-0.2cm}
	\subsection{Performance for 3D Position and Velocity Estimation}
	In the following, we evaluate the 3D position and velocity estimation performance of the proposed cooperative bistatic ISAC framework with Algorithm \ref{alg_para_esti} and \ref{alg_pos_vel_Esti}.
	In each generation of the simulation, $N_t = 2$ tBSs are randomly selected from the 8 available BSs. 
	Then, $N_{\mathrm{r}}$ rBSs are randomly chosen from the remaining 6 BSs. 
	All selected tBSs transmit the sensing signals with equal power.
	Due to the absence of established data fusion algorithms for multi-target cooperative sensing within multi-antenna BS networks, we adopt the position estimation baseline using the averaging-based hard fusion approach in \eqref{equation:average_position}, termed \textbf{Averaging}.
	
	
	\begin{figure}[tbp]
		\centering
		\begin{subfigure}{0.48\columnwidth}
			\centering
			\includegraphics[width=\linewidth]{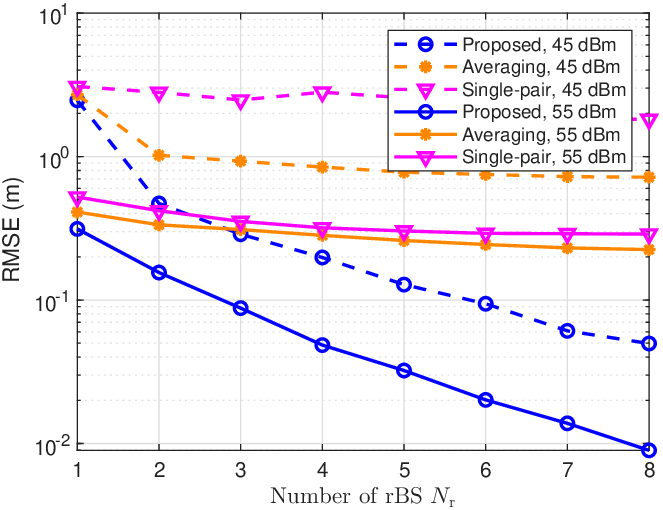}
			\caption{3D Position}
			\label{fig:RMSE_position3D_vsBSnum}
		\end{subfigure}
		\hfill 
		\begin{subfigure}{0.48\columnwidth}
			\centering
			\includegraphics[width=\linewidth]{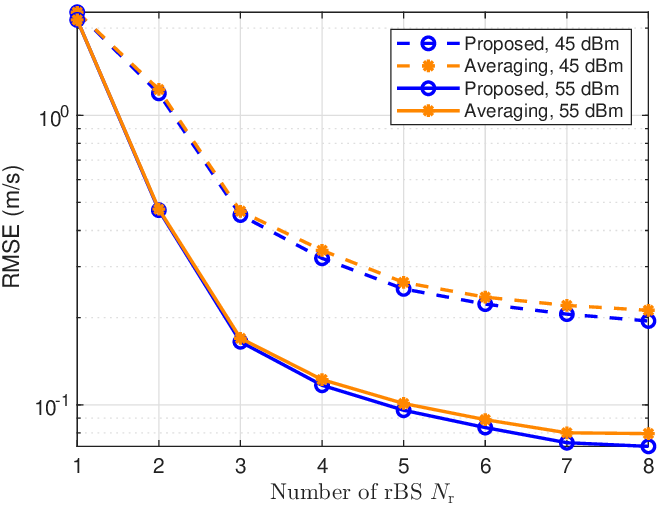}
			\caption{3D Velocity}
			\label{fig:RMSE_velocity3D_vsBSnum}
		\end{subfigure}
		
		\caption{Cooperative sensing accuracy versus the number of rBSs $N_{\mathrm{r}}$.} 
		\vspace{-0.2cm}
		\label{fig:vsBSnum}
	\end{figure}
	
	Fig. \ref{fig:vsBSnum} shows the 3D position and velocity estimation RMSE performance of both schemes under the scenario with $K=3$ targets. 
	A scheme termed \textbf{Single-pair} is also introduced as the localization performance baseline.	
	For any target, this scheme selects the basic position estimate generated by its nearest tBS and rBS as the optimal result.
	Therefore, its performance represents the upper bound of localization performance for non-cooperative bistatic sensing schemes under the considered scenario.
	The results in Fig. \ref{fig:RMSE_position3D_vsBSnum} reveal that significant accuracy improvements are achieved through multi-BS cooperation at identical transmit power levels (45 dBm or 55 dBm).
	Even with the minimum configuration ($N{\mathrm{t}}\times N{\mathrm{r}}=2$ tBS-rBS pairs) and the Averaging scheme, cooperative localization outperforms its non-cooperative counterpart.
	In addition, the results in Fig. \ref{fig:vsBSnum} shows that the proposed scheme substantially enhances multi-target localization accuracy compared to the baselines, thereby enabling higher 3D velocity estimation accuracy.
	This performance gain becomes increasingly significant with the addition of cooperative BSs, highlighting the advantages of our proposed soft fusion algorithm over traditional hard fusion methods.
	
	\begin{figure}[tbp]
		\centering
		\begin{subfigure}{0.48\columnwidth}
			\centering
			\includegraphics[width=\linewidth]{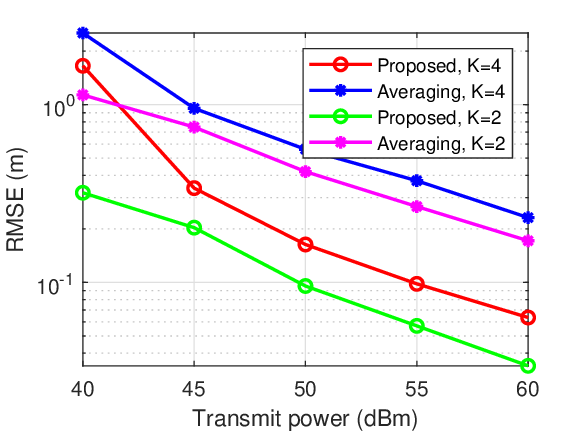}
			\caption{3D Position}
			\label{fig:RMSE_position3D_vsPt}
		\end{subfigure}
		\hfill 
		\begin{subfigure}{0.48\columnwidth}
			\centering
			\includegraphics[width=\linewidth]{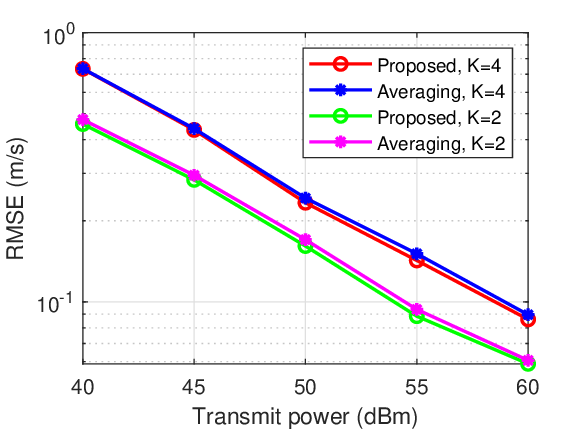}
			\caption{3D Velocity}
			\label{fig:RMSE_velocity3D_vsPt}
		\end{subfigure}
		
		\caption{Cooperative sensing accuracy versus transmit power.} 
		\vspace{-0.2cm}
		\label{fig:vsPt}
	\end{figure}
	
	Fig. \ref{fig:vsPt} examines the impact of tBS transmit power on the RMSE of 3D position and velocity estimation. 
	As expected, the RMSE monotonically decreases with increasing transmit power. 
	This trend aligns with the enhanced parameter estimation accuracy observed in Fig. \ref{fig:Para_extract}, where higher SNR improve sensing parameter extraction.
	Furthermore, Fig. \ref{fig:vsPt} consistently demonstrates the superior positioning and velocity estimation accuracy achieved by the proposed data fusion scheme compared to the baseline. 
	The performance gap remains significant throughout all power levels.

	\section{Conclusion}\label{Sec_Conclusion}
	This paper proposed a cooperative bistatic ISAC framework for LAE applications, enabling high-accuracy multi-target sensing through standardized 5G NR infrastructure. 
	By leveraging CP tensor decomposition with Vandermonde structural constraints, we developed a low-complexity algorithm for joint estimation of bistatic ranges, Doppler velocities, and AoAs from the multi-dimensional received echoes. 
	To resolve data association ambiguity across distributed transceivers, an MST-based fusion scheme was designed, eliminating outliers and enabling robust 3D position and velocity reconstruction. 
	Simulation results validated millimetre-level range accuracy and decimetre-level localization, demonstrating superior scalability for dense UAV scenarios.

	%
	\bibliographystyle{IEEEtran}
	\bibliography{IEEEabrv,ref}

\begin{thebibliography}{10}
\providecommand{\url}[1]{#1}
\csname url@samestyle\endcsname
\providecommand{\newblock}{\relax}
\providecommand{\bibinfo}[2]{#2}
\providecommand{\BIBentrySTDinterwordspacing}{\spaceskip=0pt\relax}
\providecommand{\BIBentryALTinterwordstretchfactor}{4}
\providecommand{\BIBentryALTinterwordspacing}{\spaceskip=\fontdimen2\font plus
\BIBentryALTinterwordstretchfactor\fontdimen3\font minus
  \fontdimen4\font\relax}
\providecommand{\BIBforeignlanguage}[2]{{%
\expandafter\ifx\csname l@#1\endcsname\relax
\typeout{** WARNING: IEEEtran.bst: No hyphenation pattern has been}%
\typeout{** loaded for the language `#1'. Using the pattern for}%
\typeout{** the default language instead.}%
\else
\language=\csname l@#1\endcsname
\fi
#2}}
\providecommand{\BIBdecl}{\relax}
\BIBdecl

\bibitem{9737357}
F.~Liu, Y.~Cui, C.~Masouros, J.~Xu, T.~X. Han, Y.~C. Eldar, and S.~Buzzi,
  ``Integrated sensing and communications: Toward dual-functional wireless
  networks for 6g and beyond,'' \emph{IEEE Journal on Selected Areas in
  Communications}, vol.~40, no.~6, pp. 1728--1767, Jun. 2022.

\bibitem{8999605}
F.~Liu, C.~Masouros, A.~P. Petropulu, H.~Griffiths, and L.~Hanzo, ``Joint radar
  and communication design: Applications, state-of-the-art, and the road
  ahead,'' \emph{IEEE Trans. Commun.}, vol.~68, no.~6, pp. 3834--3862, Jun.
  2020.

\bibitem{9456851}
Q.~Wu, J.~Xu, Y.~Zeng, D.~W.~K. Ng, N.~Al-Dhahir, R.~Schober, and A.~L.
  Swindlehurst, ``A comprehensive overview on {5G}-and-beyond networks with
  {UAVs}: From communications to sensing and intelligence,'' \emph{IEEE J. Sel.
  Areas Commun.}, vol.~39, no.~10, pp. 2912--2945, Oct. 2021.

\bibitem{9367457}
L.~Liu and S.~Zhang, ``A two-stage radar sensing approach based on
  {MIMO}-{OFDM} technology,'' in \emph{Proc. IEEE Globecom Workshops (GC
  Wkshps)}, Dec. 2020, pp. 1--6.

\bibitem{9724260}
L.~Pucci, E.~Paolini, and A.~Giorgetti, ``System-level analysis of joint
  sensing and communication based on 5g new radio,'' \emph{IEEE J. Sel. Areas
  Commun.}, vol.~40, no.~7, pp. 2043--2055, Jul. 2022.

\bibitem{9860521}
L.~Leyva, D.~Castanheira, A.~Silva, and A.~Gameiro, ``Two-stage estimation
  algorithm based on interleaved {OFDM} for a cooperative bistatic {ISAC}
  scenario,'' in \emph{Proc. IEEE Veh. Technol. Conf. (VTC)}, Jun. 2022, pp.
  1--6.

\bibitem{WAN2020}
X.~Wan, J.~Yi, W.~Zhan, D.~Xie, K.~Shu, J.~Song, F.~Cheng, Y.~Rao, Z.~Gong, and
  H.~Ke, ``Research progress and development trend of the
  multi-illuminator-based passive radar,'' \emph{J. Radars}, vol.~9, no.~6, pp.
  939--958, Dec. 2020.

\bibitem{10614082}
Z.~Wei, H.~Liu, Z.~Feng, H.~Wu, F.~Liu, Q.~Zhang, and Y.~Du, ``Deep cooperation
  in isac system: Resource, node and infrastructure perspectives,'' \emph{IEEE
  Internet Things Mag.}, vol.~7, no.~6, pp. 118--125, Nov. 2024.

\bibitem{10273396}
Z.~Wei, W.~Jiang, Z.~Feng, H.~Wu, N.~Zhang, K.~Han, R.~Xu, and P.~Zhang,
  ``Integrated sensing and communication enabled multiple base stations
  cooperative sensing towards {6G},'' \emph{IEEE Netw.}, vol.~38, no.~4, pp.
  207--215, Jul. 2024.

\bibitem{10032141}
P.~Gao, L.~Lian, and J.~Yu, ``Cooperative {ISAC} with direct localization and
  rate-splitting multiple access communication: A pareto optimization
  framework,'' \emph{IEEE J. Sel. Areas Commun}, vol.~41, no.~5, pp.
  1496--1515, May 2023.

\bibitem{zhenkunzhang}
Z.~Zhang, H.~Ren, C.~Pan, S.~Hong, D.~Wang, J.~Wang, and X.~You, ``Target
  localization in cooperative isac systems: A scheme based on {5G} {NR} {OFDM}
  signals,'' \emph{IEEE Trans. Commun.}, vol.~73, no.~5, pp. 3562 -- 3578, May
  2025.

\bibitem{9724258}
Q.~Shi, L.~Liu, S.~Zhang, and S.~Cui, ``Device-free sensing in {OFDM} cellular
  network,'' \emph{IEEE J. Sel. Areas Commun.}, vol.~40, no.~6, pp. 1838--1853,
  Jun. 2022.

\bibitem{10615952}
X.~Lu, Z.~Wei, R.~Xu, L.~Wang, B.~Lu, and J.~Piao, ``Integrated sensing and
  communication enabled multiple base stations cooperative {UAV} detection,''
  in \emph{Proc. IEEE Int. Conf. on Commun. Workshops (ICC Workshops)}, Aug.
  2024, pp. 1882--1887.

\bibitem{10226276}
Z.~Wei, R.~Xu, Z.~Feng, H.~Wu, N.~Zhang, W.~Jiang, and X.~Yang, ``Symbol-level
  integrated sensing and communication enabled multiple base stations
  cooperative sensing,'' \emph{IEEE Trans. Veh. Technol.}, vol.~73, no.~1, pp.
  724--738, Jan. 2024.

\bibitem{10787076}
Z.~Wei, H.~Liu, H.~Li, W.~Jiang, Z.~Feng, H.~Wu, and P.~Zhang, ``Integrated
  sensing and communication enabled cooperative passive sensing using mobile
  communication system,'' \emph{IEEE Trans. Mob. Comput.}, Early Access, Dec.
  2024, doi:{\color{blue}10.1109/TMC.2024.3514113}.

\bibitem{Han2024}
Z.~Han, H.~Ding, L.~Han, L.~Ma, X.~Zhang, M.~Lou, Y.~Wang, J.~Jin, Q.~Wang,
  G.~Liu, and J.~Wang, ``Cellular network based multistatic integrated sensing
  and communication systems,'' \emph{IET Commun.}, vol.~18, no.~20, pp.
  1878--1888, Dec. 2024.

\bibitem{10616023}
W.~Jiang, Z.~Wei, S.~Yang, Z.~Feng, and P.~Zhang, ``Cooperation-based joint
  active and passive sensing with asynchronous transceivers for perceptive
  mobile networks,'' \emph{IEEE Trans. Wireless Commun.}, vol.~23, no.~10, pp.
  15\,627--15\,641, Jul. 2024.

\bibitem{JunTang}
J.~Tang, Y.~Yu, C.~Pan, H.~Ren, D.~Wang, J.~Wang, and X.~You, ``Cooperative
  {ISAC}-empowered low-altitude economy,'' \emph{IEEE Trans. Wireless Commun.},
  vol.~24, no.~5, pp. 3837--3853, May 2025.

\bibitem{zhouzhou}
Z.~Zhou, J.~Fang, L.~Yang, H.~Li, Z.~Chen, and R.~S. Blum, ``Low-rank tensor
  decomposition-aided channel estimation for millimeter wave {MIMO}-{OFDM}
  systems,'' \emph{IEEE J. Sel. Areas Commun.}, vol.~35, no.~7, pp. 1524--1538,
  Jul. 2017.

\bibitem{11069254}
R.~Wang, H.~Ren, C.~Pan, R.~Weng, G.~Zhou, and J.~Wang, ``Channel estimation
  for {mmWave} high-mobility systems with {5G} new radio {OFDM},'' \emph{IEEE
  Trans. Commun.}, vol.~73, no.~11, pp. 11\,291--11\,307, Nov. 2025.

\bibitem{ruoyu_zhang}
R.~Zhang, L.~Cheng, S.~Wang, Y.~Lou, Y.~Gao, W.~Wu, and D.~W.~K. Ng,
  ``Integrated sensing and communication with massive {MIMO}: A unified tensor
  approach for channel and target parameter estimation,'' \emph{IEEE Trans.
  Wireless Commun.}, vol.~23, no.~8, pp. 8571--8587, Aug. 2024.

\bibitem{2024arXiv241220349R}
\BIBentryALTinterwordspacing
Z.~{Ren}, C.~{Pan}, H.~{Ren}, D.~{Wang}, L.~{Xu}, and J.~{Wang},
  ``{Two-Timescale Design for {AP} Mode Selection of Cooperative {ISAC}
  Networks},'' 2024. [Online]. Available:
  \url{https://arxiv.org/abs/2412.20349}
\BIBentrySTDinterwordspacing

\bibitem{3GPP138213}
\BIBentryALTinterwordspacing
3GPP, ``{5G}; {NR}; physical layer procedures for control,'' {3rd Generation
  Partnership Project (3GPP)}, Technical Specification (TS) 38.213, Oct. 2023,
  version 17.7.0. [Online]. Available:
  \url{https://www.etsi.org/deliver/etsi_ts/138200_138299/138213/}
\BIBentrySTDinterwordspacing

\bibitem{3GPP138104}
\BIBentryALTinterwordspacing
------, ``{5G}; {NR}; base station ({BS}) radio transmission and reception,''
  {3rd Generation Partnership Project (3GPP)}, Technical Specification (TS)
  38.104, Oct. 2025, version 19.2.0. [Online]. Available:
  \url{https://www.etsi.org/deliver/etsi_ts/138100_138199/138104/}
\BIBentrySTDinterwordspacing

\bibitem{GaoffTensor}
J.~Du, M.~Han, Y.~Chen, L.~Jin, H.~Wu, and F.~Gao, ``Tensor decompositions for
  integrated sensing and communications,'' \emph{IEEE Commun. Mag.}, vol.~62,
  no.~9, pp. 128--134, Sep. 2024.

\bibitem{C-DRCNN}
J.~Du, M.~He, J.~He, J.~Liu, L.~Jin, and Y.~Guan, ``A tensor-based signal
  processing for {ISAC} using {C-DRCNN} in {RIS}-assisted mmwave {MIMO}-{OFDM}
  systems,'' \emph{IEEE Internet Things J.}, vol.~11, no.~18, pp.
  29\,470--29\,485, Sep. 2024.

\bibitem{timevarying_TVT}
J.~Wang, W.~Zhang, Y.~Chen, Z.~Liu, J.~Sun, and C.-X. Wang, ``Time-varying
  channel estimation scheme for uplink {MU}-{MIMO} in {6G} systems,''
  \emph{IEEE Trans. Veh. Technol.}, vol.~71, no.~11, pp. 11\,820--11\,831, Nov.
  2022.

\bibitem{zxd}
X.~Zhang, \emph{Matrix analysis and applications}.\hskip 1em plus 0.5em minus
  0.4em\relax Beijing, CHN: Tsinghua University Press, 2004.

\bibitem{AlwinOnKruskal}
A.~Stegeman and N.~D. Sidiropoulos, ``On kruskal’s uniqueness condition for
  the candecomp/parafac decomposition,'' \emph{Linear Algebra Appl.}, vol. 420,
  no.~2, pp. 540--552, 2007.

\bibitem{VanLoan1993}
\BIBentryALTinterwordspacing
C.~F. Van~Loan and N.~Pitsianis, \emph{Approximation with Kronecker
  Products}.\hskip 1em plus 0.5em minus 0.4em\relax Springer Netherlands, 1993,
  pp. 293--314. [Online]. Available:
  \url{https://doi.org/10.1007/978-94-015-8196-7_17}
\BIBentrySTDinterwordspacing

\bibitem{eckart1936approximation}
C.~Eckart and G.~Young, ``The approximation of one matrix by another of lower
  rank,'' \emph{Psychometrika}, vol.~1, no.~3, pp. 211--218, 1936.

\bibitem{KRUSKAL197795}
J.~B. Kruskal, ``Three-way arrays: rank and uniqueness of trilinear
  decompositions, with application to arithmetic complexity and statistics,''
  \emph{Linear Algebra Appl.}, vol.~18, no.~2, pp. 95--138, 1977.

\bibitem{blind_signal}
M.~Sørensen and L.~De~Lathauwer, ``Blind signal separation via tensor
  decomposition with vandermonde factor: Canonical polyadic decomposition,''
  \emph{IEEE Trans. Signal Process.}, vol.~61, no.~22, pp. 5507--5519, Nov.
  2013.

\bibitem{9583869}
X.~Zhang, F.~Wang, and H.~Li, ``An efficient method for cooperative
  multi-target localization in automotive radar,'' \emph{IEEE Signal Proc.
  Let.}, vol.~29, pp. 16--20, Oct. 2022.

\bibitem{PrimShortest}
R.~C. Prim, ``Shortest connection networks and some generalizations,''
  \emph{Bell Syst. Tech. J.}, vol.~36, no.~6, pp. 1389--1401, Nov. 1957.

\bibitem{KruskalOn}
J.~B. Kruskal, ``On the shortest spanning subtree of a graph and the traveling
  salesman problem,'' in \emph{Proc. Am. Math. Sot.}, vol.~7, no.~1, 1956, pp.
  48--50.

\bibitem{GillQuasi}
P.~E. Gill and W.~Murray, ``Quasi-newton methods for unconstrained
  optimization,'' \emph{IMA J. Appl. Math.}, vol.~9, no.~1, pp. 91--108, Feb.
  1972.

\bibitem{3GPP13810101}
\BIBentryALTinterwordspacing
3GPP, ``{5G}; {NR}; user equipment ({UE}) radio transmission and reception;
  part 1: Range 1 stand alone,'' {3rd Generation Partnership Project (3GPP)},
  Technical Specification (TS) 38.101-1, Aug. 2025, version 18.10.0. [Online].
  Available: \url{https://www.etsi.org/deliver/etsi_ts/138100_138199/13810101/}
\BIBentrySTDinterwordspacing

\bibitem{ITURRECP525}
\BIBentryALTinterwordspacing
ITU, ``Calculation of free-space attenuation,'' {International
  Telecommunication Union (ITU)}, Recommendation (R) {P.525-4}, Aug. 2019.
  [Online]. Available: \url{https://www.itu.int/rec/R-REC-P.525}
\BIBentrySTDinterwordspacing

\end{thebibliography}

\end{document}